\documentclass[10pt,journal,compsoc]{IEEEtran}
\usepackage{amsmath,amsfonts}
\usepackage{algorithmic}
\usepackage{algorithm}
\usepackage{array}
\usepackage[caption=false,font=normalsize,labelfont=sf,textfont=sf]{subfig}
\usepackage{textcomp}
\usepackage{stfloats}
\usepackage{url}
\usepackage{verbatim}
\usepackage{cite}
\usepackage{capt-of}
\usepackage[utf8]{inputenc}
\usepackage{graphicx}
\usepackage{booktabs} 
\usepackage{multirow}
\usepackage{tabularx}
\usepackage{enumitem}
\usepackage[hidelinks]{hyperref}
\usepackage{orcidlink}
\usepackage{threeparttable}

\begin{document}

\title{Hot Wire 5D+: Evaluating Cognitive and Motor Trade-offs of Visual Feedback for 5D Augmented Reality Trajectories}

\author{Christian~Masuhr\orcidlink{0000-0002-9081-7844},
Julian~Koch\orcidlink{0000-0003-3700-6425},
Arne~Wendt\orcidlink{0000-0002-5782-3468},
and~Thorsten~Schüppstuhl\orcidlink{0000-0002-9616-3976}
\IEEEcompsocitemizethanks{\IEEEcompsocthanksitem C. Masuhr, J. Koch, A. Wendt and T. Schüppstuhl are with the Institute of Aircraft Production Technology, Hamburg University of Technology (TUHH), Hamburg, Germany.\\}% <-this % stops a space
}

\markboth{Preprint Version}%
{Shell \MakeLowercase{\textit{et al.}}: A Sample Article Using IEEEtran.cls for IEEE Journals}

% --- 
\IEEEpubid{\begin{minipage}{\textwidth}\ \\[12pt]
This work has been submitted to the IEEE for possible publication. Copyright may be transferred without notice, after which this version may no longer be accessible.
\end{minipage}}
% ---

\makeatletter
\let\@oldmaketitle\@maketitle
\renewcommand{\@maketitle}{\@oldmaketitle
  \begin{center}
    \setcounter{figure}{0}
    \includegraphics[width=1\linewidth]{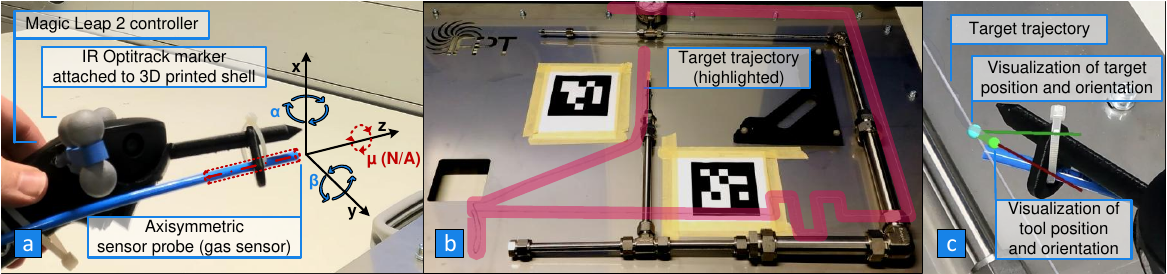}
    \captionof{figure}{Overview of the experimental setup for the 5D+ trajectory task. \textbf{(a)} The custom-built inspection tool integrating a Magic Leap 2 controller, a simulated gas axisymmetric sensor probe, and infrared markers for external OptiTrack validation. \textbf{(b)} The physical piping system mock-up with target trajectory. \textit{Note: Originally grey target trajectory has been superimposed in purple for illustrative purposes.} \textbf{(c)} Egocentric view demonstrating the continuous 5D+ visual feedback provided to the user during the tracking task.}
    \label{fig1:leakage_inspection}
  \end{center}
}
\makeatother

\maketitle

\begin{abstract}
Augmented Reality (AR) is increasingly utilized to guide users through complex spatial tasks in domains such as manufacturing, non-destructive testing, and surgery. These applications often require strict compliance with 5D+ trajectories using rotation-symmetric tools (3D position, 2D orientation, and movement speed). However, the sensori-motor baselines of untrained users during these multidimensional tracing tasks, along with the cognitive-motor trade-offs induced by varying visual feedback paradigms, remain underexplored. We present a controlled within-subjects user study (N=30) evaluating three distinct AR UI concepts for trajectory guidance, both with and without explicit orientation constraints. We analyzed spatial, orientational, and speed compliance based on the internal AR tracking, which was validated against a high-precision external optical tracking system to rule out hardware drift. By segmenting the execution into transient and steady-state phases and applying Aligned Rank Transform (ART) ANOVA, we isolated the interaction effects between visual design and task complexity. Alongside subjective metrics (NASA-TLX, SUS), our results establish conservative performance baselines for novice users performing freehand 5D trajectory following. We reveal orientation-induced cognitive-motor trade-offs and identify mitigating UI synergies. Ultimately, we provide empirical baselines and actionable design guidelines for developing effective AR guidance systems.
\end{abstract}


\section{Introduction}

\IEEEPARstart{T}{he} integration of Augmented Reality (AR) has become pivotal for manual tool guidance in critical domains like surgery, manufacturing, and maintenance \cite{dargan_augmented_2023}. Wearable optical see-through head-mounted displays (HMDs) offer hands-free, in-situ visual feedback while simultaneously digitizing complex workflows via continuous tool-tracking. By delivering real-time spatial guidance, AR mitigates the risks of error-prone manual tasks, enhances procedural accuracy, and enables objective performance evaluation.

Traditional AR guidance predominantly focuses on discrete point-to-point 3D navigation or full 6-degree-of-freedom (6-DoF) object manipulation. However, many critical operations (e.g., welding \cite{doshi2017}, drilling \cite{schoop2016}, or medical insertions \cite{sun2020}) demand continuous trajectory tracing using axisymmetric tools where rotation around the tool's longitudinal axis is functionally irrelevant. Eliminating this redundant sixth dimension preserves cognitive capacity for managing a critical, yet frequently neglected constraint: continuous movement speed. We conceptualize this specific demand for precise 5D spatial alignment strictly coupled with continuous speed control as a ``5D+'' trajectory task. Mastering this dynamic is paramount for reliable execution in continuous spatial operations, such as probe-based industrial gas leak detection (Fig. \ref{fig1:leakage_inspection}).

While foundational work has explored 6-DoF tracking \cite{zhai1997} and velocity-centric AR teleoperation \cite{su2021}, the synthesis into a cohesive 5D+ mid-air task remains under-explored. Existing approaches often rely on physical anchors (e.g., a welding torch resting on a workpiece) \cite{su2021, ceyssens2024}, which mechanically mitigate the well-documented challenges of anisotropic depth perception in AR \cite{zhai1997}. In contrast, fully mid-air tasks lack this mechanical support. Users must maintain strict spatial alignment and constant velocity entirely in free space, relying solely on visual AR cues and proprioception.

Consequently, there is a critical lack of validated visual metaphors capable of simultaneously communicating 5-DoF spatial deviations alongside speed errors without inducing severe visual clutter and cognitive overload \cite{li2024usability}. Furthermore, standard aggregate metrics (e.g., overall RMSE for the task) fail to capture dynamic behavioral adjustments in continuous pursuit-tracking \cite{bottcher2024track}. To address these gaps, we present a controlled, within-subjects user study ($N=30$) evaluating human performance in a complex 5D+ mid-air trajectory task across three distinct AR visualization concepts. The core contributions of this work are as follows:

\begin{itemize}
    \item \textbf{Empirical Baselines for 5D+ Operations:} We quantify the  sensori-motor capabilities of untrained users and the limitations for continuous 5D (3D position + 2D orientation) spatial alignment coupled with strict speed control in mid-air AR tasks.
    \item \textbf{Quantification of Cognitive-Motor Trade-offs:} By segmenting user performance into transient reaction phases and steady-state execution, we demonstrate how enforcing explicit orientation constraints induces information overload, paradoxically deteriorating continuous motor accuracy.
    \item \textbf{Actionable Design Guidelines:} We identify critical UI design synergies and provide evidence-based recommendations to balance spatial precision with cognitive demand, optimizing future AR guidance systems for high-precision spatial tasks.
\end{itemize}

To achieve these contributions, the experimental task required users to continuously trace a complex spatial trajectory, conceptually analogous to a multidimensional ''hot/buzz wire`` game. By logging AR tracking data validated against a high precision external optical tracker to rule out hardware drift and applying continuous phase segmented metrics, we successfully isolate transient cognitive reactions from steady state motor performance.

\section{Related Work}
AR-guided spatial tasks intersect human motor control, cognitive psychology, and 3D UI design. While foundational research established tracking baselines, current literature evaluates how visual metaphors impact performance and cognitive load. We categorize this work into spatial pose alignment, continuous speed control, and trajectory evaluation methods.

\subsection{Spatial Pose Alignment and the Cost of Dimensionality}
Research consistently shows that increasing the degrees of freedom (DoF) heavily impacts both human motor performance and cognitive load. Sukan et al. \cite{sukan2014} demonstrated that enforcing strict 6-DoF poses exponentially increases task completion times. Furthermore, visualizing continuous multi-axis deviations forces users to constantly shift visual attention, which fails to significantly reduce overall cognitive workload \cite{li2024usability}. While decoupling translation and rotation can improve spatial accuracy, it maintains a high physical demand during iterative adjustments \cite{benmahdjoub2025}. These multi-dimensional control challenges are exacerbated by inherent perceptual bottlenecks, particularly the significantly larger tracking errors along the depth axis caused by limits in human depth perception \cite{zhai1997}. 

To assist spatial alignment, systems like LightGuide \cite{sodhi2012} and adaptive widgets \cite{andersen2019} provide effective mid-air cues. A substantial body of this research originates from medical docking tasks, demonstrating the efficacy of AR overlays for rigid surgical insertions \cite{eom2022, condino2020} or drilling \cite{li_accuracy_2024, li2026usability}. While abstract peripheral widgets effectively reduce visual clutter and often outperform highly realistic overlapping 3D holograms during such static targeting \cite{mewes2019, seitel2011, wolf2023}, recent systematic reviews of Tool-to-Target Positioning (TOTTA) widgets emphasize the high efficacy of pre-attentive visual collimation for 5-DoF spatial alignment \cite{dastan2024totta}. Specifically, Gestalt-driven augmented collimator widgets have been shown to significantly reduce cognitive load by segregating target criteria into discrete visual gauges \cite{dastan2022,dastan2025testbedyi}. However, these widget taxonomies have been predominantly validated in static or semi-static pose matching scenarios. Moving beyond static applications to continuous surface tracing introduces distinct temporal dynamics \cite{Matsumaru2017, fischer2025}. Extrapolating the efficacy of explicitly separated widgets—which inherently demand repeated saccadic eye movements—to fully dynamic, continuous 5D+ tracing tasks remains highly questionable.

\subsection{Continuous Speed Control and the Mid-Air Challenge}
While spatial alignment is widely researched, the temporal dimension of continuous AR guidance is frequently neglected. Joos et al. \cite{joos2023} noted the difficulties of maintaining precise continuous inputs without tactile support. Addressing speed directly, Ceyssens et al. \cite{ceyssens2023, ceyssens2024} found that moving ghost visualizations effectively regulate translational speed. Similarly, Su et al. \cite{su2021} presented velocity-centric motion mapping for robotic tele-welding.

These advanced trajectory studies inherently afforded implicit haptic support, such as a welding torch resting on a workpiece \cite{su2021, ceyssens2024}. This potential mechanical support acts as a haptic anchor, which drastically mitigates the anisotropic depth perception errors inherent to AR HMDs \cite{zhai1997}. In fully unconstrained mid-air environments, the lack of such physical constraints causes users to consistently overestimate stroke lengths and exhibit significant directional deviations \cite{yi_examining_2023}. Since dynamic continuous tracking imposes cognitive demands distinct from static pose matching \cite{zhou2026}, the efficacy of simultaneously regulating visual speed and multi-axis spatial orientation in unconstrained free space remains unverified.

\subsection{Methodological Approaches to Trajectory Evaluation}
Evaluating continuous tasks requires robust tracking and precise metrics. As established by Yu et al. \cite{yu_design_2024}, XR evaluations must differentiate between discrete posture matching and continuous feedforward strategies. While spatial tracking often relies on Absolute Trajectory and Relative Pose Error \cite{zhang_tutorial_2018}, modern AR systems achieve high precision across various platforms \cite{hu2024, masuhr2025, condino2020}. 

However, relying solely on aggregated spatial metrics like overall RMSE obscures dynamic behavioral adjustments \cite{bottcher2024track}. In human motor control, goal-directed movements consist of a transient adaptation state and a steady state for stable execution \cite{bazzi_robustness_2020}. Investigating the timing of AR guidance, Ceyssens et al. \cite{ceyssens2024} demonstrated that strict live guidance inherently forces users into a reactive lag, resulting in unavoidable initial speed overshoots before trajectory acquisition. Because existing evaluations rarely segment data into these distinct phases, distinguishing such interface-induced cognitive delays from actual motor execution failures remains a significant challenge.

\section{System Design and Hypotheses}
To systematically evaluate the cognitive and motor trade-offs during continuous 5D+ trajectory tracing in mid-air, we engineered a custom AR experimental environment. The system design is structured along two primary evaluation dimensions: the conceptual visual guidance strategy (Visualization) and the presence of explicit rotational constraints (Orientation Feedback).

\subsection{General Design Rationale: Pre-attentive Color Coding}
A core element shared across all implemented interfaces is the use of dynamic color coding. As highlighted by Dastan et al. \cite{dastan2022}, color is a pre-attentive attribute that allows users to detect deviations almost instantaneously without conscious cognitive effort. By implementing a continuous color gradient from green (high precision) to red (significant error), we provide an intuitive feedback loop that supports rapid, ballistic motor corrections. This approach is particularly effective in high-stakes environments where interpreting numerical values would impose excessive cognitive load \cite{dastan2024codesign}.

\subsection{Implemented UI Concepts (Factor 1)}

To reflect real-world interactions with cohesive interfaces, we deliberately evaluate holistic HCI paradigms rather than conducting a component-level ablation study, since the visual elements inevitably influence each other. We developed three distinct UI concepts representing fundamental design philosophies (implicit-integrated, explicit-separated, and minimalist-multimodal) varying in information density and feedback integration. Their specific 5D+ technical implementations are illustrated in Figures \ref{fig:V1}, \ref{fig:V2}, and \ref{fig:V3}.

\subsubsection{V1: Tracer (Implicit Integrated Philosophy)}
\begin{figure}[!t]
\centering
\includegraphics[width=\linewidth]{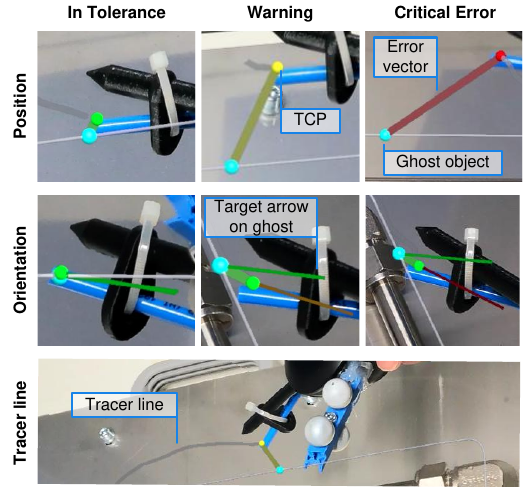}
\caption{UI Concept V1 (Tracer). Rows show positional feedback via the ghost object and added directional arrows for 5D constraints. Columns represent feedback states: In Tolerance, Warning, and Critical Error.}
\label{fig:V1}
\end{figure}

This high-immersion concept focuses on minimizing saccadic eye movements by co-locating all spatial and temporal feedback directly at the interaction site (Figure \ref{fig:V1}). The design synthesizes the findings of Ceyssens et al. \cite{ceyssens2023}, utilizing a virtual Ghost object as a pacer to regulate translational speed. By integrating the TCP color and a Target arrow directly into this moving metaphor, the interface aims to provide a continuous, intuitive "flow" that reduces the need for analytical processing during the tracing task.

\subsubsection{V2: Amplified Gestalt (Explicit Separated Philosophy)}
\begin{figure}[!t]
\centering
\includegraphics[width=\linewidth]{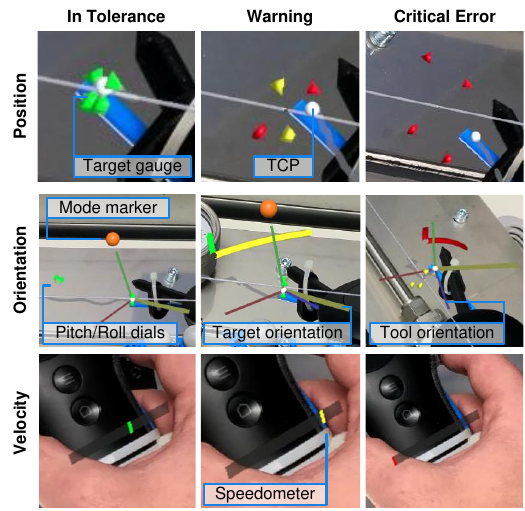}
\caption{UI Concept V2 (Gestalt). Rows illustrate the separated target gauge, pitch/roll dials for 5D constraints, and the linear speedometer. Columns represent the three feedback states.}
\label{fig:V2}
\end{figure}

V2 is grounded in Gestalt principles, such as reification and closure, to aid complex multi-axis orientation. Previous studies demonstrated that users can more effectively seek collimation when errors are separated into distinct geometric widgets \cite{dastan2022, dastan2024totta}. Consequently, this concept utilizes a static target gauge and explicit pitch/roll dials (Figure \ref{fig:V2}). By separating the components, we aim to prevent the visual clutter often caused by overlapping 3D holograms \cite{wolf2023} and provide a clearer deterministic representation of sub-millimeter deviations via an analytical speedometer.

\subsubsection{V3: Reduced (Minimalist Multimodal Philosophy)}
\begin{figure}[!b] 
\centering
\includegraphics[width=\linewidth]{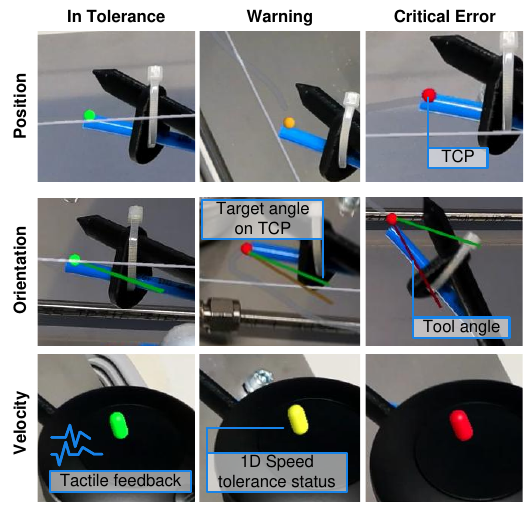}
\caption{UI Concept V3 (Reduced). Rows depict the minimalist TCP dot, abstract tool angle, and controller LED status. Speed errors are communicated via vibrotactile feedback.}
\label{fig:V3}
\end{figure}

To address visual overload in 5D tasks, the \textit{Reduced} concept (Figure \ref{fig:V3}) offloads velocity pacing to alternative sensory pathways \cite{Yang2026}. It preserves foveal attention for primary spatial navigation through visual decluttering, displaying only a minimalist Tool Center Point (TCP) for position and a simple directional target arrow for orientation. To test the limits of sensory substitution, we combined a 1D status LED-Widget with standard frequency-modulated vibrotactile feedback on the controller. We deliberately accepted the inherent unidimensionality of standard actuators: this scalar feedback signals the presence and intensity of a velocity error, but cannot encode its directional vector (i.e., moving too fast vs. too slow). Consequently, V3 serves as an experimental boundary condition to investigate whether the cognitive relief of extreme visual decluttering can outweigh the usability costs of ambiguous, non-directional tactile pacing.

\subsection{The Orientation Factor (Factor 2)}
The second independent variable manipulates task complexity by enforcing or hiding the orientation widgets described above.
\begin{itemize}
    \item \textbf{Task A (Without Orientation):} Focuses solely on 3D position and speed to establish a foundational tracking baseline.
    \item \textbf{Task B (With Orientation):} Enforces the full 5D+ constraint by activating the angular visualizers to evaluate the cognitive-motor trade-offs of multi-axis alignment.
\end{itemize}

\subsection{Research Questions and Hypotheses}
To systematically investigate human capabilities and limitations during 5D+ trajectory tracing, this study is guided by four primary Research Questions (RQs):

\begin{description}[leftmargin=0pt, labelindent=0pt, itemsep=4pt, parsep=0pt]
    \item[RQ1 (Effectiveness and Load):] How do the three distinct UI concepts compare regarding objective tracking performance and subjective cognitive load?
    
    \item[RQ2 (The Cost of Orientation):] How does the introduction of a secondary orientation alignment task impact the primary positional and speed performance as well as the overall cognitive demand?
    
    \item[RQ3 (Synergy and Usability):] Which UI concept provides the best overall system usability and most effectively mitigates the cognitive-motor costs induced by the orientation constraints?
    
    \item[RQ4 (Temporal Dynamics):] How do the user's spatial and speed compliance differ between the initial transient phase and the continuous steady-state phase across the different UI concepts?
\end{description}

\noindent Based on established principles of perceptual psychology, visual clutter, and sensory substitution, we formulated the following testable hypotheses:

\begin{description}[leftmargin=0pt, labelindent=0pt, itemsep=4pt, parsep=0pt]
    \item[\textbf{H1a (Spatial and Orientation Performance):}] The explicit component separation of \textit{V2 (Amplified Gestalt)} will yield the highest precision, resulting in significantly lower position and orientation errors compared to \textit{V1} and \textit{V3}.
    
    \item[\textbf{H1b (Speed Control):}] The immersive ghost metaphor of \textit{V1 (Tracer)} will provide the most effective speed pacing, resulting in significantly lower speed errors compared to the explicit gauges of \textit{V2} and the tactile feedback of \textit{V3}.
    
    \item[\textbf{H2 (Cognitive-Motor Trade-off):}] Enforcing explicit 5D orientation constraints will significantly improve orientation compliance but deteriorate position and speed accuracy, while simultaneously increasing cognitive load across all visual conditions.
    
    \item[\textbf{H3 (Subjective Load and Usability):}] The minimalist multimodal approach of \textit{V3 (Reduced)} will effectively offload visual demand and induce the lowest cognitive load. Conversely, the high information density of \textit{V2} will induce visual clutter, leading to the highest cognitive load and lowest usability scores.
    
    \item[\textbf{H4 (Phase Dynamics):}] Transient phases will exhibit significantly higher spatial and speed errors than steady-state phases. We anticipate that \textit{V2}'s explicit feedback enables faster stabilization, whereas \textit{V1} suffers from prolonged initial overshoots.
\end{description}

\section{Methodology}
The study design isolates the cognitive and motor effects of distinct AR visualization strategies during a continuous 5D+ mid-air trajectory task.

\subsection{Participants}
We recruited 30 participants ($N=30$; 5 female, 25 male; mean age $M = 30.5$ years, $SD = 9.25$). The cohort was recruited from a university environment, consisting primarily of undergraduate students and academic researchers. While this demographic does not encompass professional domain experts, it provides an ideal sample for evaluating fundamental sensori-motor baselines without the confounds of use-case-specific motor routines. All participants had normal or corrected-to-normal vision (one color blindness, not affecting task performance), reported no motor impairments, and provided written informed consent.

\subsection{Apparatus and Setup}
The experimental setup simulated an unconstrained industrial inspection task in need of sub-centimeter tracking accuracy. Participants wore a Magic Leap 2 (ML2) optical see-through HMD, rendering visual feedback at 60 Hz (like shown in Figure \ref{fig1:leakage_inspection}c). The ML2 controller was embedded in a custom 3D-printed shell mimicking a gas leak sensor hose and probe (Figure \ref{fig1:leakage_inspection}a).

To validate tracking stability, we utilized the external motion capture Optitrack system with six Prime X13 cameras, providing a reported accuracy of 0.2 mm. Infrared markers on the HMD and tool enabled real-time mapping of the OptiTrack coordinate system into the Unity AR space via a Kabsch rigid-body transformation. Notably, the OptiTrack data did not replace the primary AR tracking. Instead, it served as an independent validation layer to confirm the absence of progressive spatial drift. The task was performed entirely standing, enforcing a strict mid-air execution without haptic anchors.

\subsection{Experimental Design}
\begin{figure}[!t]
        \centering
        \includegraphics[width=1\linewidth]{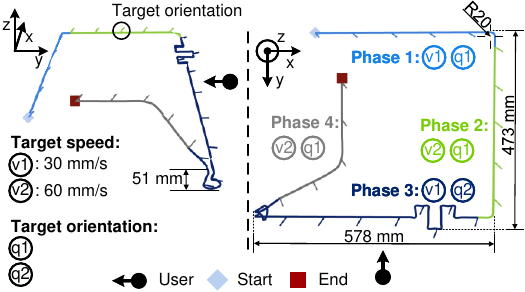}
        \caption{Overview of the target trajectory and its distinct spatial segments relative to the underlying piping structure in two different views.}
        \label{fig:trajectory}
    \end{figure}

The core task required participants to guide the sensor's Tool Center Point (TCP) along a continuous 3D trajectory overlaid on a virtual piping system (Figure \ref{fig1:leakage_inspection}b, \ref{fig:trajectory}). This simulated leak inspection scenario forces simultaneous control of position, orientation, and speed under realistic spatial occlusion. The spatial geometry combines a planar raster path for systematic surface scanning and a 3D curved element simulating a pipe fitting.

To evaluate both transient adaptation and steady execution, the trajectory is divided into four consecutive phases with distinct complexity shifts:
\textbf{Phase 1} establishes a baseline for continuous positional tracking and speed compliance.
\textbf{Phase 2} introduces an immediate speed shift to test rapid motor adaptation.
\textbf{Phase 3} introduces the highest spatial complexity by enforcing multi axis angular constraints along the curved fitting while maintaining baseline speed.
\textbf{Phase 4} demands simultaneous adaptation to new speed and orientation targets within low trajectory complexity.

The enforced orientation targets $(q_{1},q_{2})$ remained ergonomically neutral (tilted toward the user) requiring solely a discrete 90-degree roll adjustment between execution phases. This ensures the task evaluated cognitive switching rather than testing joint mobility limits. Exact system parameters and tolerances are detailed in Figure \ref{fig:trajectory}.

To eliminate stationary reaction noise, the initial trajectory data prior to consistent forward motion was truncated. We defined the transient Start phase as a fixed two second temporal window. Consequently, the geometric distance covered during this phase inherently varies based on the respective target speed, successfully capturing true physical acceleration and trajectory convergence.

The independent variables were the Visualization Concept (V1: Tracer, V2: Amplified Gestalt, V3: Reduced) and the Orientation Feedback (Task A: Position and Speed vs. Task B: Position, Speed, and Orientation), resulting in exactly six trials per participant. To mitigate potential learning and fatigue effects, the order of the evaluated visualization concepts was counterbalanced across participants using a Latin Square design.

The 45-minute procedure began with a briefing and ML2 fitting, followed by 5 minutes of training with the \textit{Tracer} UI to ensure familiarization with mid-air tracking. During the main study, participants executed the six counterbalanced trials. To ensure fresh recall, the NASA-TLX was administered immediately after each trial. The session concluded with the SUS for each concept, a preference ranking, and a qualitative interview.

\subsection{Data Processing and Metrics}

\subsubsection{Motor Variance Isolation}
\label{motor_variance}
Previous research on ML2 inside-out tracking demonstrated a baseline spatial accuracy of 1.39 mm \cite{masuhr2025}. However, ArUco-instantiated AR coordinate systems inherently exhibit minor static global offsets upon initialization. Because users react exclusively to the visual overlay rendered within the HMD and naturally compensate for these initial system disparities, evaluating the Absolute Trajectory Error (ATE) would falsely attribute this setup noise to human motor failure \cite{sodhi2012}.

Therefore, in line with established trajectory evaluation methodologies \cite{zhang_tutorial_2018, sodhi2012}, we calculated the Relative Pose Error (RPE) to strictly isolate localized motor variance. External OptiTrack validation confirmed that the AR tracking remained highly stable during task execution, without progressive mid-trial drift. We applied a two-pass Trimmed Iterative Closest Point (TrICP) algorithm \cite{fischer2025} to compute the optimal rigid-body transformation between the recorded AR trajectory and the target path. This robust alignment neutralizes initial global offsets and minimizes the influence of extreme tracking outliers, ensuring the reported metrics exclusively represent true human motor variance.

\subsubsection{Filtering and Phase Segmentation}
Raw AR logs were transformed into a local reference frame. A low-pass Butterworth filter (4th order, 6 Hz cutoff) mitigated high-frequency tracking noise. While this inherently produce time domain overshoots, it ensures robust extraction of the underlying macro-motor control behavior. To separate early adjustment behavior from stable control, each sample was assigned to the nearest reference trajectory index and split into two temporal segments: a \textit{Transient State} (labeled as ''Start``; the first two seconds after phase onset) and a \textit{Steady State} (labeled as ''Run``; stable execution). To prevent intra-trial autocorrelation, global main and interaction effects were evaluated on length-adjusted, trial-aggregated data ($df=115$), while temporal dynamics utilized the phase-segmented data.

\subsubsection{Performance and Subjective Metrics}
To strictly isolate spatial deviations from temporal velocity errors, the positional accuracy ($RMSE_{Pos}$) was calculated based on the shortest Euclidean distance between the measured tool positions and a high resolution interpolated reference trajectory utilizing a KD tree search. Orientation compliance ($RMSE_{Ori}$) was computed as the root mean square of the angular deviations between the measured forward vectors and their respective reference forward vectors. Continuous speed control ($RMSE_{Speed}$) was calculated as the root mean square deviation between the absolute spatial velocity and the target speed. Velocity was derived directly from the 3D distance between consecutive samples rather than projecting vectors onto the reference path. This deliberately includes compensatory zig zag movements to accurately reflect the true physical tip speed, which is critical for industrial scanning.

The perceived subjective workload was evaluated over six dimensions using a 5 point Task Load Index (NASA TLX) administered after each trial condition. Furthermore, overall system acceptability was assessed once per visualization modality using the standard 10 item System Usability Scale (SUS) on a 5 point Likert scale.

\section{Results}
Due to non-parametric tracking data, primary inferential statistics utilize the ART ANOVA \cite{wobbrock2011}. To model repeated measures and accommodate marginal design imbalances, analyses employ Linear Mixed Models (LMM) with participant as a random effect. Post-hoc pairwise comparisons use Estimated Marginal Means (EMMs) with Holm-Bonferroni corrections ($p_{adj}$). Confirming successful counterbalancing, an initial ART ANOVA found no significant block order effects on positional error ($F(5, 139.6) = 1.65, p = .152, \eta_p^2 = .056$). Comprehensive results and methodological rationale are detailed in the supplemental materials.

\begin{figure}[!t]
    \centering
    \includegraphics[width=1\linewidth]{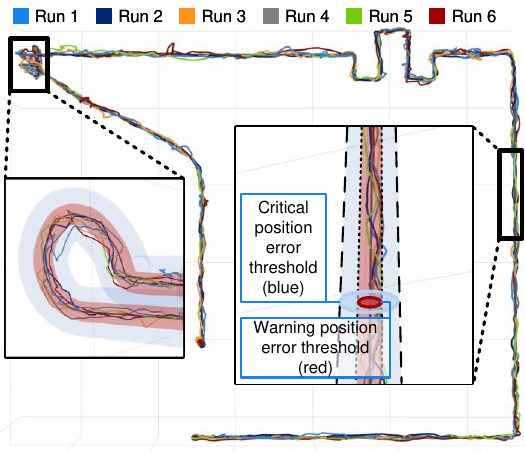}
    \caption{Example of a recorded trajectory path from a user and the tilt axis for the plane used for the depth error estimation.}
    \label{fig:trajectory_overview}
\end{figure}

\begin{figure*}[!t]
    \centering
    \includegraphics[width=1\textwidth]{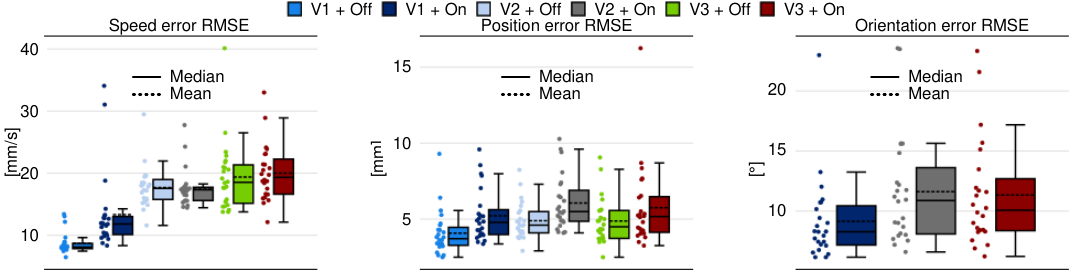}
    \caption{Distribution plots of the main error metrics: Speed Error (A), Position Error (B), and Orientation Error (C) with orientation guidelines (+ On) enabled and disabled (+ Off). The mean and median RMSE is visualized for each condition.}
    \label{fig:error_metrics}
\end{figure*}

\subsection{Spatial Accuracy (Position Error)}
To evaluate overall spatial precision, we analyzed the root mean square error for position ($RMSE_{Pos}$). The ART ANOVA revealed a highly significant main effect of Visualization on spatial accuracy ($F(2, 115) = 12.39, p < .001, \eta_p^2 = .177$). Post-hoc Holm-corrected comparisons indicated the highly immersive \textit{Tracer} UI achieved significantly lower positional errors compared to both \textit{Amplified Gestalt} ($p < .001$) and \textit{Reduced} UIs ($p = .007$). \textit{Gestalt} and \textit{Reduced} showed no significant difference ($p = .053$).

This indicates that the added physical challenge of managing tool rotation universally degrades positional precision, and none of the evaluated UI concepts could completely shield users from this baseline spatial drift.

%This indicates that the added physical challenge of managing tool rotation universally degrades positional precision, and none of the evaluated UI concepts could completely shield users from this baseline spatial drift.

Furthermore, enforcing 5D constraints systematically deteriorated spatial alignment (Orientation effect: $F(1, 115) = 31.99, p < .001, \eta_p^2 = .218$). The Visualization $\times$ Orientation interaction was not significant ($F(2, 115) = 1.05, p = .354$).

\subsection{Continuous Speed Control}
The global analysis showed a strong main effect of Visualization on speed error ($F(2, 115) = 67.89, p < .001, \eta_p^2 = .541$). Aligning with our hypotheses, \textit{Tracer} provided highly effective pacing, significantly outperforming \textit{Gestalt} ($p < .001$) and \textit{Reduced} ($p < .001$), with no difference between the baselines ($p = .263$). While Orientation had no general main effect ($p = .205$), a significant Visualization $\times$ Orientation interaction emerged ($F(2, 115) = 4.03, p = .021, \eta_p^2 = .065$). Post-hoc analysis uncovered a distinct trade-off: \textit{Tracer}'s speed control significantly deteriorated when 5D constraints were enforced ($p = .002$). Conversely, both \textit{Gestalt} and \textit{Reduced} successfully buffered this cognitive load, showing no significant speed degradation (both $p = .472$).

\subsection{Phase Dynamics and Transient Effects}
Segmented temporal analysis ($RMSE_{Speed}$) isolated early adaptation from stable execution. The ART ANOVA yielded a highly significant Visualization $\times$ Segment interaction ($F(14, 1363.1) = 6.91, p < .001, \eta_p^2 = .066$).

Contrast testing confirmed significant transient effects (overshoots) during sudden acceleration (Phase 2 Start vs. Phase 2 Run; Figure \ref{fig:speed_dynamics}). The speed error for the \textit{Tracer} concept was significantly higher during the initial adaptation phase compared to the subsequent steady state ($p = .0072, d = 0.110$). In contrast, the \textit{Gestalt} and \textit{Reduced} UIs exhibited no significant transient shock, maintaining stable performance during the initial acceleration ($p = 1.000$).

Furthermore, analyzing the stable execution phases (steady states) revealed fundamental differences in how the UIs handle shifting track geometries. When transitioning from a phase with low geometric complexity (Phase 2 Run) to the segment with the most complex geometric shapes (Phase 3 Run), the \textit{Tracer} maintained an extraordinarily stable error profile with no significant performance shift ($p = 1.000, d = 0.026$). The \textit{Reduced} UI also remained relatively stable ($p = .676, d = 0.071$). Conversely, the \textit{Gestalt} UI exhibited a significant performance shift between these steady states ($p < .001, d = 0.206$). 

Finally, when moving from the highly complex Phase 3 into the geometrically less complex final segment (Phase 4 Run), speed error significantly increased across all concepts ($p < .001$), potentially indicating cumulative fatigue or terminal loss of focus. However, the \textit{Reduced} UI showed the smallest relative performance drop ($d = 0.183$) compared to the \textit{Tracer} ($d = 0.250$) and \textit{Gestalt} ($d = 0.311$) interfaces.

\subsection{The Cost of Orientation}
Analyzing angular error ($RMSE_{Ori}$) exclusively for Task B (Orientation On) trials, the one-way ART ANOVA showed a significant main effect of the Visualization concept on orientation compliance ($F(2, 665) = 16.06, p < .001, \eta_p^2 = .046$). Contrary to initial expectations, Holm-corrected post-hoc contrasts revealed that the highly immersive \textit{Tracer} UI achieved significantly lower angular deviations compared to both the explicitly separated \textit{Gestalt} UI ($p < .001, r = .16$) and the \textit{Reduced} UI ($p < .001, r = .20$). There was no significant difference in angular accuracy between the \textit{Gestalt} and \textit{Reduced} concepts ($p = .244, r = .05$).

\subsection{Cognitive Load and Usability}
Subjective workload was first evaluated using the aggregated NASA-TLX Overall score (Table \ref{tab:synthesis_subjective_objective}). The ART ANOVA revealed significant main effects for both the Visualization concept ($F(2, 142.0) = 36.49, p < .001, \eta_p^2 = .339$) and Orientation guidance ($F(1, 143.2) = 50.15, p < .001, \eta_p^2 = .259$). However, the overall interaction effect between UI and Orientation was not significant ($p = .238$). 

To isolate the specific cognitive costs, we subsequently analyzed the NASA-TLX Mental Load subscale. Here, alongside the expected main effects, a highly significant Visualization $\times$ Orientation interaction emerged ($F(2, 142.1) = 5.97, p = .003, \eta_p^2 = .077$). Post-hoc analyses uncovered a distinct cognitive shielding effect: While transitioning to Task B (Orientation On) caused a significant surge in mental demand for both the \textit{Gestalt} ($p < .001, d = 0.46$) and \textit{Reduced} ($p = .002, d = 0.31$) interfaces, the \textit{Tracer} UI fully buffered this cognitive spike, yielding no significant increase in perceived mental load ($p = .143, d = 0.16$).

System Usability Scale (SUS) scores differed highly significantly across the tested conditions ($F(2, 58) = 33.36, p < .001, \eta_p^2 = .535$). The \textit{Tracer} UI achieved an excellent usability rating ($M = 80.58 \pm 14.69$), significantly outperforming the \textit{Reduced} UI ($p < .001, d = 0.55$) and the \textit{Gestalt} UI ($p < .001, d = 1.07$).

\subsection{Directional Error Decomposition (Planar vs. Depth)}
Decomposing the spatial errors revealed a consistent directional bias: deviation in depth along the standing user's viewing axis ($RMSE_Y$) was notably larger than errors within the perpendicular viewing plane ($RMSE_{XZ}$). The depth error accounted for up to 55.4\% of the total spatial deviation (Figure \ref{fig:trajectory_overview} and  Table \ref{tab:synthesis_subjective_objective}). 

The ART ANOVA revealed significant main effects of the Visualization concept for both planar ($F(2, 139.0) = 14.67, p < .001, \eta_p^2 = .174$) and depth dimensions ($F(2, 139.0) = 12.88, p < .001, \eta_p^2 = .156$), with the Tracer UI significantly outperforming all other concepts (all $p < .01$). Furthermore, the \textit{Gestalt} UI exhibited significantly higher planar error than the \textit{Reduced} UI ($p = .013, d = 0.214$).

The Tracer demonstrated superior resilience by buffering the cognitive load of orientation constraints (Task B). Unlike the susceptible Gestalt and Reduced UIs ($p = .508$), its depth error proportion remained stable ($51.9\% \rightarrow 51.4\%$), suggesting integrated feedback preserves spatial precision despite secondary motor demands.

\begin{table*}[!t]
\centering
\caption{Summary of trial-aggregated ART ANOVA omnibus results and selected post-hoc contrasts with effect sizes.}
\label{tab:master_stats}
\setlength{\tabcolsep}{4pt} % tabularx profitiert von etwas Padding
\renewcommand{\arraystretch}{1.05}
\footnotesize
\begin{tabularx}{\textwidth}{@{}
  >{\raggedright\arraybackslash}X 
  >{\raggedright\arraybackslash}p{2.6cm}
  c c c 
  | 
  >{\raggedright\arraybackslash}p{2.8cm}
  c c c
  @{}}
& \multicolumn{4}{c}{\textbf{Omnibus Test (Main and Interaction Effects)}} & \multicolumn{4}{c}{\textbf{Holm-Corrected Post-Hoc Contrasts}} \\
\cmidrule(lr){2-5} \cmidrule(l){6-9}
\textbf{Metric} & \textbf{Effect} & \textbf{\textit{F}-value} & \textbf{\textit{p}-value} & \textbf{$\eta_p^2$} & \textbf{Contrast} & \textbf{\textit{t}-ratio} & \textbf{\textit{p}\textsubscript{adj}} & \textbf{Size ($d$)} \\
\midrule
\multirow{3}{*}{\textbf{Spatial accuracy} ($RMSE_{Pos}$)}
& Visualization (UI) & 12.39 & \textbf{$< .001$} & 0.177 & Gestalt -- Tracer & 4.942 & \textbf{$< .001$} & 0.46 \\
& Orientation        & 31.99 & \textbf{$< .001$} & 0.218 & Reduced -- Tracer & 2.987 & \textbf{$.007$} & 0.28 \\
& UI $\times$ Orientation & 1.05 & $.354$ & 0.018 & Gestalt -- Reduced & 1.955 & $.053$ & 0.18 \\
\midrule
\multirow{4}{*}{\textbf{Speed control} ($RMSE_{Speed}$)}
& Visualization (UI) & 67.89 & \textbf{$< .001$} & 0.541 & Gestalt -- Tracer & 10.024 & \textbf{$< .001$} & 0.93 \\
& Orientation        & 1.62 & $.205$ & 0.014 & Reduced -- Tracer & 9.481 & \textbf{$< .001$} & 0.88 \\
& \multirow{2}{*}{UI $\times$ Orientation} & \multirow{2}{*}{4.03} & \multirow{2}{*}{\textbf{$.021$}} & \multirow{2}{*}{0.065} & Tracer (Off -- On) & -3.529 & \textbf{$.002$} & 0.55 \\
& & & & & Gestalt (Off -- On) & -1.121 & $.472$ & 0.18 \\
\midrule
\multirow{2}{*}{\textbf{Orientation error} ($RMSE_{Ori}$)}
& \multirow{2}{*}{Visualization (UI)} & \multirow{2}{*}{13.59} & \multirow{2}{*}{\textbf{$< .001$}} & \multirow{2}{*}{0.319} & Gestalt -- Tracer & 4.435 & \textbf{$< .001$} & 0.82 \\
& & & & & Reduced -- Tracer & 3.611 & \textbf{$.002$} & 0.67 \\
\bottomrule
\multicolumn{9}{@{}p{\textwidth}@{}}{\footnotesize \textit{Note:} Omnibus tests evaluate trial-aggregated metrics ($df_{res} = 115$; Orientation $df_{res} = 58$) to prevent intra-trial pseudoreplication. Effect size for omnibus tests is $\eta_p^2$; for contrasts, Cohen's $d$ is used.}
\end{tabularx}
\end{table*}latin

\subsection{Global Baselines and Local Trajectory Dynamics}
Contextualizing the inferential statistics requires examining absolute error magnitudes (Figure \ref{fig:error_metrics}, Table \ref{tab:synthesis_subjective_objective}). The average positional deviation remained below one centimeter across all conditions, ranging from $M = 4.1$\,mm (\textit{Tracer}, Task A) to $M = 6.1$\,mm (\textit{Gestalt}, Task B). Temporal errors revealed a wider absolute performance gap: the \textit{Tracer} ($M = 8.7$\,mm/s in Task A) more than halved the speed error compared to baseline interfaces (e.g., \textit{Reduced}: $M = 19.4$\,mm/s). Finally, enforcing 5D constraints (Task B) introduced substantial angular deviations. However, the \textit{Tracer} maintained the highest compliance ($M = 9.16^\circ$), outperforming both \textit{Reduced} ($M = 11.34^\circ$) and \textit{Gestalt} ($M = 11.63^\circ$).

\begin{figure}[!t]
    \centering
    \includegraphics[width=1\linewidth]{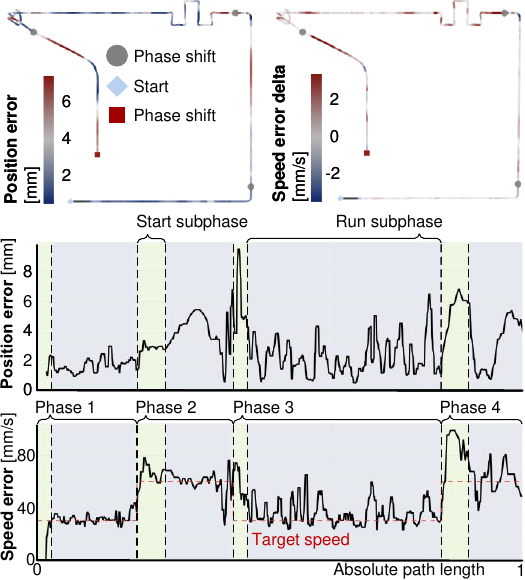}
    \caption{Temporal dynamics and speed error distribution along the evaluation trajectory of a representative single-user trial using the \textit{Tracer} UI with orientation widgets activated. \textbf{Top:} A 3D spatial heatmap illustrating localized position and speed deviations. \textbf{Bottom:} Continuous speed profile over the normalized absolute path length highlighting transient speed overshoots and positional errors during initial acceleration (Start phases) before settling into steady-state execution (Run phases).}
    \label{fig:speed_dynamics} 
\end{figure}

Beyond aggregated baselines, 3D trajectory heatmaps (Figure \ref{fig:speed_dynamics}) reveal that spatial errors were not uniformly distributed. While the \textit{Tracer} achieved the highest global spatial accuracy, visual inspection highlights localized positional spikes during complex geometric transitions. Conversely, the \textit{Gestalt} UI, despite its higher global error baseline, maintained a visually more consistent spatial error envelope during these sharp curves. This observation perfectly aligns with the transient temporal overshoots of the \textit{Tracer} identified in Section 5.3 (Figure \ref{fig:speed_dynamics}).

Furthermore, continuous tracking logs revealed that these spatial error envelopes are characterized by high-frequency micro-oscillations (alternating over- and undershooting) typical for closed-loop visuomotor corrections.

\subsection{Qualitative Feedback and Thematic Analysis}
To contextualize the objective performance metrics, a single researcher conducted an exploratory, inductive thematic analysis of the open-ended survey responses. Following a data-driven approach, free-text comments were iteratively coded and grouped into overarching themes to identify recurring usability patterns

\textbf{V1: Tracer (Implicit Guidance \& Mimicry):}
The \textit{Tracer} concept was predominantly praised for centralizing visual attention. Twelve participants ($n=12$) explicitly noted that aggregating positional, temporal, and orientation errors into a single moving target enabled a ''mimicry-based`` strategy rather than requiring mental calculation. Participants described this guidance as providing ''cognitive relief`` (P9). Regarding the 5D constraints, users frequently reported that the integrated visualization felt intuitive to follow. However, four participants ($n=4$) reported issues with visual occlusion, noting that the ghost marker could physically obscure the actual tool tip during perfect execution. Furthermore, participants noted that spatial depth estimation remains difficult when relying solely on implicit 3D cues (P4).

\textbf{V2: Amplified Gestalt (Visual Fragmentation):}
While the peripheral gauges were appreciated by a minority ($n=4$) for providing precise raw metrics without requiring head movements, they induced the most frequent negative theme: the foveal-peripheral split ($n=15$). Participants reported that the spatial separation between the tooltip and the peripheral speedometer forced constant, exhausting saccades, summarized that as ''too large to monitor simultaneously`` from a user (P22).

\textbf{V3: Reduced (Feedback Ambiguity):}
The minimalist approach successfully reduced eye strain for some users ($n=5$), and the spatial tolerance anchors were highlighted as effective compensations for limited stereoscopic depth ($n=5$). However, a critical design flaw regarding feedback unidimensionality was identified by half of the participants ($n=14$). While encoding error severity, the purely scalar haptic vibration lacked directional cues, forcing frequent ''corrective guessing`` about speed adjustments (P27).

\section{Discussion}
Our findings provide critical insights into the cognitive-motor bottlenecks of continuous 5D mid-air trajectory tracing. While some established AR design paradigms hold true for static tasks, our empirical data reveals a  divergent dynamic between visual attention, motor execution, and perceived workload during continuous movement.

\begin{table*}[!htbp]
\centering
\caption{Synthesis of Subjective Perception, Objective Performance, and Directional Error Decomposition across UI Concepts. Values are presented as Mean (SD). Arrows indicate whether higher ($\uparrow$) or lower ($\downarrow$) values represent better outcomes.}
\label{tab:synthesis_subjective_objective}
\footnotesize
\resizebox{\textwidth}{!}{%
\begin{tabular}{@{}ll | cccc | cccc | cc@{}}
\toprule
& & \multicolumn{4}{c|}{\textbf{Subjective Perception}} & \multicolumn{4}{c|}{\textbf{Spatial Error (mm) ($\downarrow$)}} & \multicolumn{2}{c}{\textbf{Other Errors ($\downarrow$)}} \\
\cmidrule(lr){3-6} \cmidrule(lr){7-10} \cmidrule(l){11-12}
\textbf{UI Concept} & \textbf{Orientation} & 
\begin{tabular}{@{}c@{}}\textbf{SUS ($\uparrow$)} \\ \textbf{Score}\end{tabular} & 
\begin{tabular}{@{}c@{}}\textbf{TLX ($\downarrow$)} \\ \textbf{Overall}\end{tabular} & 
\begin{tabular}{@{}c@{}}\textbf{TLX ($\downarrow$)} \\ \textbf{Mental}\end{tabular} & 
\begin{tabular}{@{}c@{}}\textbf{TLX ($\downarrow$)} \\ \textbf{Success}\end{tabular} & 
\begin{tabular}{@{}c@{}}\textbf{Total} \\ \textbf{($RMSE_{Pos}$)}\end{tabular} & 
\begin{tabular}{@{}c@{}}\textbf{Planar} \\ \textbf{($RMSE_{XZ}$)}\end{tabular} & 
\begin{tabular}{@{}c@{}}\textbf{Depth} \\ \textbf{($RMSE_Y$)}\end{tabular} & 
\begin{tabular}{@{}c@{}}\textbf{Depth} \\ \textbf{(\%)}\end{tabular} & 
\begin{tabular}{@{}c@{}}\textbf{Speed} \\ \textbf{(mm/s)}\end{tabular} & 
\begin{tabular}{@{}c@{}}\textbf{Orientation} \\ \textbf{($^\circ$)}\end{tabular} \\
\midrule
\multirow{2}{*}{Gestalt} & Off & 44.8 (14.83) & 2.98 (0.81) & 3.47 (1.31) & 3.23 (1.07) & 4.9 (1.2) & 3.3 (0.9) & 3.5 (0.9) & 51.9\% & 17.7 (3.3) & N/A \\
                         & On  & 44.8 (14.83) & 3.74 (0.82) & 5.28 (1.36) & 4.48 (1.38) & 6.1 (1.8) & 4.0 (1.3) & 4.4 (1.2) & 52.7\% & 17.6 (3.0) & 11.63 (4.60) \\
\midrule
\multirow{2}{*}{Reduced} & Off & 62.0 (21.45) & 2.82 (0.61) & 3.03 (1.19) & 3.70 (1.42) & 4.9 (1.5) & 3.0 (0.8) & 3.7 (1.3) & 55.4\% & 19.4 (5.6) & N/A \\
                         & On  & 62.0 (21.45) & 3.30 (0.98) & 4.21 (1.63) & 4.45 (1.30) & 5.8 (2.7) & 3.7 (1.9) & 4.3 (2.0) & 53.7\% & 20.0 (4.6) & 11.34 (4.45) \\
\midrule
\multirow{2}{*}{Tracer}  & Off & 80.6 (14.69) & 2.19 (0.70) & 1.97 (1.10) & 2.53 (1.28) & 4.1 (1.5) & 2.7 (1.0) & 3.0 (1.2) & 51.9\% & 8.7 (1.7) & N/A \\
                         & On  & 80.6 (14.69) & 2.79 (0.77) & 2.52 (1.18) & 3.28 (1.03) & 5.2 (1.7) & 3.6 (1.1) & 3.7 (1.4) & 51.4\% & 13.3 (6.3) & 9.16 (3.51) \\
\bottomrule
\end{tabular}%
}
\end{table*}

\subsection{Continuous Tracing (H1) and Experimental Tactile Offloading (H3)}

Contrary to previous literature favoring separated widgets for spatial alignments \cite{dastan2022}, hypothesis H1a is firmly rejected. The highly immersive \textit{Tracer} UI achieved significantly lower positional and orientational errors compared to both baseline concepts. Concurrently, hypothesis H1b is accepted, as the \textit{Tracer} significantly outperformed all other concepts in speed compliance, effectively mastering all 5D+ constraints simultaneously.

Two core factors explain this superiority. First, the UIs enforce fundamentally different manual control paradigms: the \textit{Gestalt} UI functions as a compensatory tracking display (visualizing system error as corrective feedback), while the \textit{Tracer} facilitates pursuit tracking (visualizing target and tool independently as continuous feedforward \cite{yu_design_2024}). While providing effective foundation for static docking \cite{dastan2022}, the spatial decoupling of compensatory widgets can become counterproductive in dynamic tasks, causing a foveal-peripheral split that disrupts tight sensori-motor control loops. Second, the \textit{Tracer} inherently couples spatial and temporal constraints. Matching a moving 3D target intrinsically dictates the correct speed, naturally minimizing temporal deviations.

Regarding H3, the results present a dual picture. As expected, the high information density of the \textit{Gestalt} UI (\textit{V2}) induced the highest cognitive load and lowest usability (Table \ref{tab:synthesis_subjective_objective}). However, tactile offloading in the \textit{Reduced} UI (\textit{V3}) failed to lower cognitive demand. Although \textit{V3} matched \textit{Gestalt} in objective speed control ($p = .097$), it provided no subjective relief. Qualitative feedback ($n=14$) revealed that 1D haptic vibrations lacking a directional vector force exhausting ''corrective guessing``. This ambiguity aligns with the subjective metrics (Table \ref{tab:synthesis_subjective_objective}): users rated their task success under tactile pacing significantly worse than with the \textit{Tracer}, reducing \textit{V3}'s overall usability scores.

\subsection{The Cognitive-Motor Trade-off and Illusion of Competence (H2)}
Supporting H2, enforcing orientation constraints creates a 5D bottleneck, confirming the observations of Dastan et al. \cite{dastan2022}. Crucially, the AR visualization dictates whether this manifests as subjective cognitive strain or objective motor degradation.

The explicitly separated \textit{Gestalt} UI absorbed the 5D complexity cognitively. Enforcing orientation caused a massive surge in perceived mental demand (Section 5.5). However, this heightened focus preserved temporal precision, preventing any significant degradation in speed control (Section 5.2). 

Conversely, the \textit{Tracer} UI absorbed the constraint within the motor domain, fostering an \textit{Illusion of Competence}. Lacking explicit gauges, it successfully buffered the cognitive spike, yielding no significant increase in mental load. However, this self-assessed comfort masked a significant deterioration in objective speed control ($p < .001$). Although the \textit{Tracer} retained the best absolute performance, its implicit mimicry caused users to subconsciously prioritize spatial alignment over temporal pacing. This divergence is striking (Table \ref{tab:synthesis_subjective_objective}): while actual speed error worsened significantly under 5D constraints, self-rated task success (TLX Success) remained extremely positive. Users felt perfectly in control despite eroding motor precision.

This double dissociation exposes a critical risk: highly immersive AR metaphors effectively shield users from cognitive overload but obscure relative physical performance drops. Consequently, our findings highlight a critical methodological takeaway: evaluating immersive tools solely through subjective metrics like the NASA-TLX is insufficient, as perceived mastery can drastically diverge from actual execution.

\subsection{Temporal Dynamics and Phase Adaptation (H4)}
Transient \textit{Start} phases consistently exhibited significantly higher errors than stable \textit{Run} phases. Accepting H4, this confirms that AR-guided motor control is a dynamic process requiring continuous recalibration during task shifts, rather than a static execution.

This temporal decomposition exposes a fundamental limitation of implicit guidance. Despite achieving the lowest global speed error, the \textit{Tracer} suffered a significant transient overshoot during the Phase 2 \textit{Start} acceleration (Section 5.3). This reflects a forced transition into reactive control. In pursuit tracking, humans naturally rely on proactive feed-forward mechanisms. However, because the \textit{Tracer} exactly colocates the target with the required position, it strips away the predictive horizon, forcing the user into a strictly closed-loop feedback response. This inherent delay in processing the visual delta leads to the observed motor lag and over-correction, corroborating the findings on live-guidance timing by Ceyssens et al. \cite{ceyssens2024}.

Conversely, the \textit{Tracer} demonstrated superior stability during steady-state \textit{Run} phases. Entering the geometrically complex Phase 3 \textit{Run} segment, its error profile showed no significant degradation ($p = 1.000$), whereas \textit{Gestalt} performance deteriorated significantly ($p < .001$). This contrast suggests that while peripheral indicators aid in anticipating discrete speed changes during \textit{Start} phases, integrated immersive mimicry remains substantially more robust for maintaining sensori-motor flow during continuous \textit{Run} executions.

\subsection{Absolute Precision Benchmarks and Ergonomic Trade-offs}
Our experiment establishes critical baseline benchmarks for the absolute accuracy attainable in continuous 5D mid-air tracking. As participants were untrained novices, these results provide a conservative performance threshold for AR-assisted workforce onboarding.

Empirical data demonstrates that novices can achieve sub-centimeter spatial precision even without physical haptic constraints. Supported by the \textit{Tracer} metaphor, participants maintained a global average positional deviation of 4.6\,mm. Even when fully simultaneous 5D constraints were enforced, the \textit{Tracer} sustained a highly accurate positional deviation of 5.2\,mm alongside an angular compliance of $M = 9.16^\circ$. While domain experts like surgeons or specialized welders achieve higher precision through years of motor training, this baseline confirms that optimized AR interfaces can successfully bridge the motor skill gap for macro-scale spatial tasks.

Synthesizing these benchmarks with our temporal analyses reveals distinct ergonomic trade-offs. A common assumption in AR design is that immersive 3D graphics inherently increase cognitive load due to visual clutter \cite{mewes2019, wolf2023}. Our data contradicts this premise: the visually integrated \textit{Tracer} induced the lowest absolute mental demand and the highest usability score ($M = 80.6$). Offloading complex 5D tracking into an intuitive mimicry task effectively minimized cognitive strain. 

The true ergonomic cost of this immersive approach lies in physical occlusion and transient instability (Section 6.3), not mental load. While the \textit{Tracer's} volumetric overlap successfully anchors depth perception (Section 6.6), it inevitably obscures the physical tool tip during high-precision segments. Attempting to solve this occlusion through minimalist alternatives like the \textit{Reduced} UI introduces severe psychological friction via tactile ambiguity (Section 6.1). Consequently, the optimal visualization remains context-dependent: the immersive \textit{Tracer} excels in tasks prioritizing absolute global precision and low cognitive fatigue. Conversely, if a task strictly forbids visual occlusion of the workpiece, designers must resort to peripheral or multimodal interfaces, accepting the inevitable penalty in temporal pacing and cognitive strain.

\subsection{Directional Error Decomposition and Hardware Limitations}
Spatial error decomposition reveals a pronounced directional bias along the depth axis. This asymmetry aligns with known limitations of optical see-through displays, where focal distance mismatches and vergence-accommodation conflicts inherently reduce human depth acuity.

The immersive \textit{Tracer} UI mitigated this issue significantly better than the baselines. Connecting the physical tool to the virtual target with an explicit vector arrow creates a continuous visual tether. This provides salient, co-located perspective cues, allowing users to subconsciously anchor the hologram and perceive depth variations without gaze shifts.

Conversely, this decomposition highlights the limitations of separated widgets during dynamic tasks. Counterintuitively, even though the \textit{Gestalt} UI explicitly visualized depth deviations, it failed to translate this information into superior spatial accuracy. While its peripheral indicators marginally outperformed the \textit{Reduced} UI, they could not match the \textit{Tracer}. Ultimately, the cognitive cost of shifting foveal focus from the Tool Center Point to read peripheral gauges substantially outweighs the theoretical benefit of explicitly displaying depth information.

\subsection{Suitability for Industrial Applications}

% gekürzte Fassung

Mapping our absolute benchmark results to real-world scenarios confirms the practical viability of continuous AR guidance in selected domains. The achieved positional accuracy indicates that optimized AR can enable novice workers to perform sub-centimeter spatial tasks reliably, with mean errors ranging from 4.1\,mm in the absence of explicit orientation constraints to 5.1\,mm under full 5D requirements. Lower median values suggest that the means were affected by a limited number of skewed outliers. This satisfies tolerances for macro-scale processes like volumetric scanning or manual leak detection \cite{masuhr2025}. However, while our Relative Pose Error (RPE) confirms robust human motor capabilities, establishing true industrial readiness requires addressing the operational stability of the spatial coordinate frame. Since previous investigations found no relevant tracking drift for the utilized hardware \cite{masuhr2025}, uncorrected global static offsets (Absolute Trajectory Error), as opposed to human motor limitations, remain a primary bottleneck for deployment, necessitating rigorous software-side locking mechanisms in future applications.

\subsection{Design Implications for Continuous 5D+ Guidance}
Synthesizing our empirical metrics with user feedback yields four actionable design guidelines for continuous AR-guided spatial tasks.

\textbf{Avoid Out-of-Focus Peripheral Widgets:} The severe divided attention issues of the \textit{Gestalt} UI demonstrate that traditional dashboard layouts fail in dynamic spatial tasks. Offloading gauges to the visual periphery provokes a foveal-peripheral split, disrupting tight visuo-motor control loops. Process-critical indicators must be strictly co-located within the physical end-effector's foveal region to prevent performance drops.

\textbf{Compensate Reactive Lag via Target Offsetting:} As observed during transient accelerations (Section 6.3), exact temporal colocation of the \textit{Tracer} ghost forces a reactive feedback loop, causing inevitable overshoots. Corroborating recent propositions by Ceyssens et al. \cite{ceyssens2024}, our findings strongly support introducing a spatial or temporal lead. Positioning the virtual target slightly ahead of the required trajectory effectively compensates for the dead time and dynamics of the human visuomotor control system, thereby smoothing transient motor spikes.

\textbf{Encode Vectorial Direction in Haptic Substitution:} While tactile offloading lowers visual clutter, the \textit{Reduced} UI revealed a severe usability bottleneck due to ambiguous unidimensional alerts. When utilizing haptic sensory substitution, feedback must encode the specific corrective vector rather than merely signaling binary error magnitudes. Lacking directional cues, users are forced into exhausting trial-and-error counter-movements.

\textbf{Combine Implicit Flow with Explicit Anchors:} The ideal AR guidance paradigm requires a hybrid approach. Users strongly favored the \textit{Tracer}'s intuitive mimicry for maintaining overall spatial flow and velocity pacing. However, this implicit metaphor causes visual occlusion during maximum precision segments. To mitigate this, the volumetric ghost should drive the overall process flow, dynamically augmented by minimalist explicit geometric constraints only during critical fine motor adjustments.

\subsection{Limitations and Future Work}
Given that we evaluated holistic design metaphors, the impacts of individual UI components remain intertwined. Future ablation studies must systematically vary both the spatial distance and the visual encoding of peripheral indicators to identify the exact thresholds where guidance transitions from a helpful reference to a cognitive distraction.

The intentional unidimensionality of the vibrotactile feedback in the \textit{Reduced} UI poses a methodological limitation. Providing only a scalar alert without a directional vector forced users into 'corrective guessing'. Although this successfully established an extreme boundary condition for visual offloading, future iterations must employ discriminable tactile patterns that encode bidirectional errors to eliminate artificial cognitive friction.

Furthermore, evaluating novices establishes a pure baseline to demonstrate the rapid onboarding potential of optimized AR interfaces. While prior studies observed a leveling effect where visualizations equalized novice and expert performance \cite{wolf2023}, future research must investigate if this holds true in dynamic, velocity-constrained tasks. Specifically, it remains an open question whether domain experts benefit equally from immersive mimicry, or if explicitly separated UIs paradoxically disrupt their deeply ingrained motor routines.

Finally, our experimental trajectory relied on a continuous pipe geometry. Future investigations should explore how these UI concepts scale to highly irregular spatial paths, varying scale factors, or unpredictable directional reversals. Additionally, visual inspection of the steady-state tracking revealed continuous high-frequency micro-oscillations. Although typical for closed-loop visuomotor tracking, future domain-specific studies must analyze how these implicit sweeping motions impact volumetric sensor coverage during industrial inspections.

\section{Conclusion}
Evaluating human performance and cognitive-motor trade-offs in AR-based 5D+ trajectory tracing reveals that the spatial design of visual feedback fundamentally dictates motor precision and cognitive workload. Immersive, unified \textit{Tracer} metaphors prevent divided attention and enable tight sensori-motor loops, contradicting the assumption that dense 3D graphics inherently increase cognitive strain. 

However, this subjective comfort creates a dangerous Illusion of Competence: while users feel perfectly in control, implicit guidance can mask significant temporal pacing deterioration when complex orientation constraints are enforced. Furthermore, attempting to offload cognitive strain via multimodal tactile cues fails unless the feedback explicitly encodes directional correction vectors.

Because purely implicit guidance suffers from reactive motor lags and initial overshoots, future AR systems must adopt hybrid strategies to enable proactive feed-forward control. Designers should spatially or temporally offset virtual targets to compensate for human reaction times, and augment continuous flow with explicit geometric anchors during high-precision segments.

Optimized AR interfaces enable untrained users to reliably achieve sub-centimeter spatial precision, establishing a robust baseline for AR-assisted industrial onboarding. To achieve true operational readiness, future development must focus on robust software mitigation against global tracking shifts. Ultimately, by aligning visualization paradigms with process-specific physical constraints, next-generation AR applications can effectively bridge the gap between human motor capabilities and complex spatial demands.

\section*{Acknowledgement}
%This work was supported in part by the Federal Ministry for Economic Affairs and Energy of Germany (BMWE)
%under grant number 20M2219D as part of the project iPREFER.

\vspace{2pt}
\noindent
\begin{minipage}[c]{0.59\columnwidth}
%\vspace{0pt}
This work was supported in part by the Federal Ministry for Economic Affairs and Energy of Germany (BMWE)
under grant number 20M2219D as part of the project iPREFER.
\end{minipage}
\hfill
\begin{minipage}[c]{0.39\columnwidth}
%\vspace{0pt}
\centering
\includegraphics[width=0.95\linewidth]{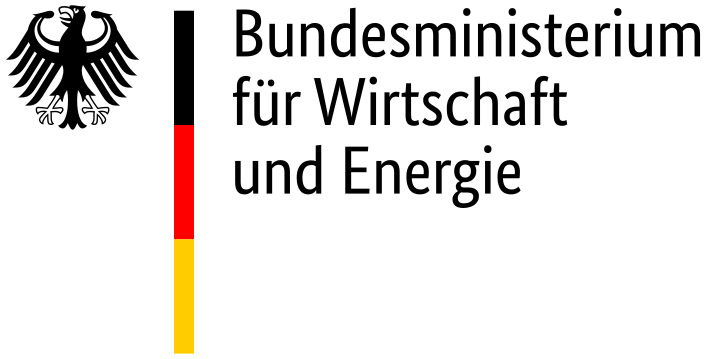}
\end{minipage}

\vspace{0.25cm}
\noindent The authors acknowledge the use of Gemini (Google) solely for language polishing and proofreading of this manuscript.

\bibliographystyle{IEEEtran}
\bibliography{Masuhr_bibliography}

\newpage

\onecolumn 

\section*{Supplemental Materials}

This supplemental document provides abbreviated and highlighted statistical outputs of our analyses. To evaluate global system efficacy overall main and interaction effects (Hypotheses 1-3) were calculated on length-adjusted, aggregated trial metrics ($df_{res} = 115$). Conversely, to isolate transient initial adaptation from steady-state motor execution (Hypothesis 4), temporal dynamics were analyzed using phase-segmented data ($df_{res} = 1381$).

\subsection*{S1. Temporal Trajectory Segments and System Thresholds}

\vspace{1cm} 

\begin{table*}[!htbp] 
    \centering
    \footnotesize
    \begin{threeparttable}
    \caption{System thresholds for the four trajectory phases dictating the visual UI states (Tolerance and Warning). While target velocity shifts to induce transient motor adaptation, spatial and angular thresholds remain constant.}
    \label{tab:trajectory_segments}
    \begin{tabular*}{\textwidth}{@{\extracolsep{\fill}} c ccc cc cc c @{}}
    \toprule
    & \multicolumn{3}{c}{\textbf{Velocity (mm/s)}} & \multicolumn{2}{c}{\textbf{Position (mm)}} & \multicolumn{2}{c}{\textbf{Angle ($^\circ$)}} & \textbf{Orientation} \\
    \cmidrule(lr){2-4} \cmidrule(lr){5-6} \cmidrule(lr){7-8} \cmidrule(l){9-9}
    \textbf{Phase} & \textbf{Target} & \textbf{Tol.} & \textbf{Warn.} & \textbf{Tol.} & \textbf{Warn.} & \textbf{Tol.} & \textbf{Warn.} & \textbf{Target State} \\
    \midrule
    \textbf{1} & $30$ & $\pm 15$ & $\pm 20$ & $\pm 4.0$ & $\pm 10.0$ & $\pm 5.0$ & $\pm 15.0$ & $q_1$ \\
    \textbf{2} & $60$ & $\pm 20$ & $\pm 30$ & $\pm 4.0$ & $\pm 10.0$ & $\pm 5.0$ & $\pm 15.0$ & $q_1$ \\
    \textbf{3} & $30$ & $\pm 15$ & $\pm 20$ & $\pm 4.0$ & $\pm 10.0$ & $\pm 5.0$ & $\pm 15.0$ & $q_2$ \\
    \textbf{4} & $60$ & $\pm 20$ & $\pm 30$ & $\pm 4.0$ & $\pm 10.0$ & $\pm 5.0$ & $\pm 15.0$ & $q_1$ \\
    \bottomrule
    \end{tabular*}
    \begin{tablenotes}[para, flushleft]
    \item \textit{Note:} All thresholds apply symmetrically. Target orientations defined as quaternions $(x,y,z,w)$: $q_1 = [-0.2826, -0.3262, -0.1028, 0.8962]$ and $q_2 = [-0.2868, 0.2868, 0.0904, 0.9096]$.
    \end{tablenotes}
    \end{threeparttable}
\end{table*}

\clearpage

\subsection*{S2. Phase Dynamics and Temporal Adaptation}

To isolate human reaction times (transient adaptation) from stable motor execution (steady state), the trajectory was divided into specific temporal segments. Table \ref{tab:detailed_descriptives} provides the uncompressed descriptive error magnitudes separated by individual tracking metrics. Unlike the global evaluation, the subsequent ART ANOVAs operate on the unaggregated phase-segmented data.

\vspace{1.0cm}

\begin{table*}[!htbp]
\centering
\footnotesize
\begin{threeparttable}
\caption{Detailed descriptive statistics of the Root Mean Square Errors (RMSE). Data is segmented into initial adaptation (Transient) and continuous execution (Steady State) phases. Values are formatted as Mean (Standard Deviation).}
\label{tab:detailed_descriptives}
\begin{tabular*}{\textwidth}{@{\extracolsep{\fill}} ll c cccccc @{}}
\toprule
 & & & \multicolumn{2}{c}{\textbf{Position Error (mm)}} & \multicolumn{2}{c}{\textbf{Speed Error (mm/s)}} & \multicolumn{2}{c}{\textbf{Orientation Error ($^\circ$)}} \\
\cmidrule(lr){4-5} \cmidrule(lr){6-7} \cmidrule(lr){8-9}
\textbf{Phase} & \textbf{UI Concept} & \textbf{Orientation} & \textbf{Transient} & \textbf{Steady State} & \textbf{Transient} & \textbf{Steady State} & \textbf{Transient} & \textbf{Steady State} \\
\midrule
\multirow{6}{*}{\textbf{Phase 1}} 
 & \multirow{2}{*}{Gestalt}   & Off & 4.4 (2.4) & 4.6 (2.0) & 22.6 (7.9) & 19.7 (6.5) & 31.15 (11.63) & 39.92 (11.53) \\
 &                            & On  & 5.1 (2.9) & 5.7 (2.9) & 16.5 (4.8) & 11.4 (4.5) &  4.66 (3.26)  &  7.59 (4.48) \\
 \cmidrule{2-9}
 & \multirow{2}{*}{Reduced}   & Off & 4.0 (3.0) & 5.3 (3.1) & 21.7 (9.5) & 16.9 (9.5) & 29.44 (13.83) & 39.45 (13.52) \\
 &                            & On  & 4.2 (2.1) & 5.7 (3.8) & 20.4 (8.8) & 16.5 (6.9) &  7.31 (5.05)  & 10.91 (6.15) \\
 \cmidrule{2-9}
 & \multirow{2}{*}{Tracer}    & Off & 4.8 (10.3)& 3.2 (1.5) & 13.4 (5.6) &  4.5 (3.7) & 26.87 (11.43) & 36.52 (11.31) \\
 &                            & On  & 4.3 (5.1) & 3.9 (2.7) & 12.6 (0.4) &  5.3 (1.9) &  5.30 (5.11)  &  5.64 (4.57) \\
\midrule
\multirow{6}{*}{\textbf{Phase 2}} 
 & \multirow{2}{*}{Gestalt}   & Off & 4.6 (1.9) & 5.3 (1.3) & 17.4 (8.1) & 19.3 (5.9) & 47.02 (12.17) & 52.32 (14.99) \\
 &                            & On  & 5.4 (2.3) & 5.8 (2.0) & 25.2 (6.2) & 22.6 (5.8) &  9.63 (7.67)  &  8.74 (7.97) \\
 \cmidrule{2-9}
 & \multirow{2}{*}{Reduced}   & Off & 4.8 (3.6) & 4.8 (1.5) & 20.0 (10.4)& 18.0 (7.8) & 47.66 (13.69) & 53.23 (16.42) \\
 &                            & On  & 4.9 (2.5) & 6.1 (6.0) & 23.3 (8.3) & 21.4 (8.1) & 12.13 (9.95)  & 12.48 (8.88) \\
 \cmidrule{2-9}
 & \multirow{2}{*}{Tracer}    & Off & 3.9 (1.1) & 4.9 (2.0) & 13.0 (3.4) &  9.5 (1.6) & 45.65 (12.75) & 51.99 (15.00) \\
 &                            & On  & 4.2 (1.9) & 5.0 (1.8) & 12.6 (2.6) &  9.6 (1.8) &  5.73 (5.57)  &  6.91 (4.51) \\
\midrule
\multirow{6}{*}{\textbf{Phase 3}} 
 & \multirow{2}{*}{Gestalt}   & Off & 3.7 (1.2) & 4.5 (1.9) & 10.6 (5.9) & 13.2 (7.8) & 21.24 (11.47) & 43.03 (12.93) \\
 &                            & On  & 6.7 (2.3) & 5.9 (2.4) & 15.9 (8.1) & 13.3 (7.9) & 48.18 (12.64) & 11.92 (8.45) \\
 \cmidrule{2-9}
 & \multirow{2}{*}{Reduced}   & Off & 4.0 (1.4) & 4.8 (2.0) & 16.4 (9.3) & 19.5 (12.7)& 23.76 (12.50) & 47.32 (13.82) \\
 &                            & On  & 6.4 (2.9) & 5.3 (1.9) & 21.0 (11.2)& 15.9 (8.4) & 39.96 (8.47)  &  8.20 (4.30) \\
 \cmidrule{2-9}
 & \multirow{2}{*}{Tracer}    & Off & 3.5 (2.9) & 3.5 (1.3) &  7.5 (4.7) &  5.9 (2.6) & 19.47 (11.29) & 44.59 (14.51) \\
 &                            & On  & 7.5 (2.7) & 4.9 (2.4) & 32.9 (13.9)& 11.7 (9.7) & 35.39 (8.04)  &  6.96 (4.50) \\
\midrule
\multirow{6}{*}{\textbf{Phase 4}} 
 & \multirow{2}{*}{Gestalt}   & Off & 4.7 (1.9) & 6.3 (1.6) & 20.6 (7.7) & 26.8 (4.6) & 26.05 (13.92) & 29.72 (13.12) \\
 &                            & On  & 7.5 (3.6) & 7.6 (2.8) & 24.3 (8.6) & 28.3 (5.7) & 49.78 (10.49) & 11.64 (10.37) \\
 \cmidrule{2-9}
 & \multirow{2}{*}{Reduced}   & Off & 5.4 (3.5) & 6.3 (2.6) & 19.2 (9.5) & 25.6 (6.2) & 26.03 (15.04) & 29.16 (14.98) \\
 &                            & On  & 6.7 (2.6) & 6.4 (2.0) & 24.5 (6.4) & 27.1 (6.3) & 46.87 (6.33)  &  9.25 (4.39) \\
 \cmidrule{2-9}
 & \multirow{2}{*}{Tracer}    & Off & 4.0 (1.5) & 6.0 (2.2) & 12.7 (2.9) & 17.7 (2.2) & 26.96 (15.05) & 28.99 (15.05) \\
 &                            & On  & 7.4 (2.9) & 6.1 (2.2) & 23.8 (12.3)& 17.4 (2.0) & 38.61 (4.95)  &  6.18 (3.19) \\
\bottomrule
\end{tabular*}
\begin{tablenotes}[para, flushleft]
\item \textit{Note:} All cell values are structured as Mean (Standard Deviation).
\end{tablenotes}
\end{threeparttable}
\end{table*}

\begin{table*}[!htbp]
\centering
\footnotesize
\begin{threeparttable}
\caption{ART ANOVA and Post-Hoc Contrasts for Spatial Phase Dynamics ($RMSE_{Pos}$).}
\label{tab:anova_phase_pos}
\begin{tabular*}{\textwidth}{@{\extracolsep{\fill}} l cccc @{}}
\toprule
\multicolumn{5}{c}{\textbf{Part 1: Omnibus ART ANOVA (Interaction Effect)}} \\
\midrule
\textbf{Effect} & \textbf{\textit{F}-value} & \textbf{\textit{df}} & \textbf{\textit{p}-value} & \textbf{Effect Size ($\eta_p^2$)} \\
\midrule
Visualization $\times$ Segment & 2.27 & 14, 1381 & \textbf{$.005$} & 0.023 \\
\midrule
\multicolumn{5}{c}{\textbf{Part 2: Selected Holm-Corrected Post-Hoc Contrasts}} \\
\midrule
\textbf{Contrast} & \textbf{Estimate} & \textbf{\textit{t}-ratio} & \textbf{\textit{p}\textsubscript{adj}} & \textbf{Effect Size ($d$)} \\
\midrule
\multicolumn{5}{l}{\textbf{\textit{A) Phase Transition: Phase 2 Start -- Phase 2 Run}}} \\
Gestalt  & -4.54 & -0.612 & $1.000$ & 0.05 \\
Reduced  & -13.00 & -1.753 & $1.000$ & 0.01 \\
Tracer   &  7.46 &  1.006 & $1.000$ & 0.08 \\
\midrule
\multicolumn{5}{l}{\textbf{\textit{B) Phase Transition: Phase 2 Run -- Phase 3 Run}}} \\
Gestalt  & 20.06 &  2.704 & $.205$  & 0.08 \\
Reduced  &  0.81 &  0.109 & $1.000$ & 0.01 \\
Tracer   & -5.84 & -0.787 & $1.000$ & 0.02 \\
\midrule
\multicolumn{5}{l}{\textbf{\textit{C) Phase Transition: Phase 3 Run -- Phase 4 Run}}} \\
Gestalt  & -33.95 & -4.577 & \textbf{$< .001$} & 0.38 \\
Reduced  & -22.75 & -3.067 & \textbf{$.060$} & 0.18 \\
Tracer   & -24.89 & -3.355 & \textbf{$.005$}   & 0.28 \\
\bottomrule
\end{tabular*}
\begin{tablenotes}[para, flushleft]
\item \textit{Note:} This ART ANOVA explicitly evaluates the interaction effect between the Visualization concept and the temporal Segment to identify concept-specific adaptation patterns.
\end{tablenotes}
\end{threeparttable}
\end{table*}

\begin{table*}[!htbp]
\centering
\footnotesize
\begin{threeparttable}
\caption{ART ANOVA and Post-Hoc Contrasts for Temporal Phase Dynamics ($RMSE_{Speed}$).}
\label{tab:anova_phase_speed}
\begin{tabular*}{\textwidth}{@{\extracolsep{\fill}} l cccc @{}}
\toprule
\multicolumn{5}{c}{\textbf{Part 1: Omnibus ART ANOVA (Interaction Effect)}} \\
\midrule
\textbf{Effect} & \textbf{\textit{F}-value} & \textbf{\textit{df}} & \textbf{\textit{p}-value} & \textbf{Effect Size ($\eta_p^2$)} \\
\midrule
Visualization $\times$ Segment & 6.91 & 14, 1363 & \textbf{$< .001$} & 0.066 \\
\midrule
\multicolumn{5}{c}{\textbf{Part 2: Selected Holm-Corrected Post-Hoc Contrasts}} \\
\midrule
\textbf{Contrast} & \textbf{Estimate} & \textbf{\textit{t}-ratio} & \textbf{\textit{p}\textsubscript{adj}} & \textbf{Effect Size ($d$)} \\
\midrule
\multicolumn{5}{l}{\textbf{\textit{A) Phase Transition: Phase 2 Start -- Phase 2 Run}}} \\
Gestalt  & -19.66 & -0.341 & $1.000$ & 0.01 \\
Reduced  &  64.22 &  1.114 & $1.000$ & 0.03 \\
Tracer   & 211.37 &  3.666 & \textbf{$.007$} & 0.11 \\
\midrule
\multicolumn{5}{l}{\textbf{\textit{B) Phase Transition: Phase 2 Run -- Phase 3 Run}}} \\
Gestalt  & 438.54 &  7.606 & \textbf{$< .001$} & 0.21 \\
Reduced  & 148.81 &  2.581 & $.675$  & 0.07 \\
Tracer   &  55.07 &  0.955 & $1.000$ & 0.03 \\
\midrule
\multicolumn{5}{l}{\textbf{\textit{C) Phase Transition: Phase 3 Run -- Phase 4 Run}}} \\
Gestalt  & -689.05 & -11.950 & \textbf{$< .001$} & 0.31 \\
Reduced  & -441.59 & -7.658  & \textbf{$< .001$} & 0.18 \\
Tracer   & -557.19 & -9.663  & \textbf{$< .001$} & 0.25 \\
\bottomrule
\end{tabular*}
\begin{tablenotes}[para, flushleft]
\item \textit{Note:} This ART ANOVA explicitly evaluates the interaction effect between the Visualization concept and the temporal Segment to identify concept-specific adaptation patterns.
\end{tablenotes}
\end{threeparttable}
\end{table*}

\begin{table*}[!htbp]
\centering
\footnotesize
\begin{threeparttable}
\caption{ART ANOVA and Post-Hoc Contrasts for Orientational Phase Dynamics ($RMSE_{Ori}$).}
\label{tab:anova_phase_ori}
\begin{tabular*}{\textwidth}{@{\extracolsep{\fill}} l cccc @{}}
\toprule
\multicolumn{5}{c}{\textbf{Part 1: Omnibus ART ANOVA (Interaction Effect)}} \\
\midrule
\textbf{Effect} & \textbf{\textit{F}-value} & \textbf{\textit{df}} & \textbf{\textit{p}-value} & \textbf{Effect Size ($\eta_p^2$)} \\
\midrule
Visualization $\times$ Segment & 2.05 & 14, 1381 & \textbf{$.013$} & 0.021 \\
\midrule
\multicolumn{5}{c}{\textbf{Part 2: Selected Holm-Corrected Post-Hoc Contrasts}} \\
\midrule
\textbf{Contrast} & \textbf{Estimate} & \textbf{\textit{t}-ratio} & \textbf{\textit{p}\textsubscript{adj}} & \textbf{Effect Size ($d$)} \\
\midrule
\multicolumn{5}{l}{\textbf{\textit{A) Phase Transition: Phase 2 Start -- Phase 2 Run}}} \\
Gestalt  & -36.96 & -1.311 & $1.000$ & 0.05 \\
Reduced  & -17.96 & -0.637 & $1.000$ & 0.02 \\
Tracer   & -32.54 & -1.155 & $1.000$ & 0.06 \\
\midrule
\multicolumn{5}{l}{\textbf{\textit{B) Phase Transition: Phase 2 Run -- Phase 3 Run}}} \\
Gestalt  & -107.42 & -3.811 & \textbf{$.002$} & 0.17 \\
Reduced  &  29.75 &  1.056 & $1.000$ & 0.05 \\
Tracer   &   4.46 &  0.158 & $1.000$ & 0.01 \\
\midrule
\multicolumn{5}{l}{\textbf{\textit{C) Phase Transition: Phase 3 Run -- Phase 4 Run}}} \\
Gestalt  & 165.75 &  5.881 & \textbf{$< .001$} & 0.31 \\
Reduced  & -21.46 & -0.761 & $1.000$ & 0.04 \\
Tracer   & -14.62 & -0.519 & $1.000$ & 0.02 \\
\bottomrule
\end{tabular*}
\begin{tablenotes}[para, flushleft]
\item \textit{Note:} This ART ANOVA explicitly evaluates the interaction effect between the Visualization concept and the temporal Segment to identify concept-specific adaptation patterns. To prevent the model from being confounded by unconstrained spatial behavior, this analysis strictly operated on a data subset containing only Task B (Orientation On) trials.
\end{tablenotes}
\end{threeparttable}
\end{table*}

\clearpage

\subsection*{S3. Global Tracking Performance (Aggregated Trial Data)}

The following tables present the comprehensive ART ANOVA results for global performance metrics. As reported in the main manuscript, these analyses operate on length-adjusted, aggregated trial averages to ensure statistical robustness. 

\vspace{1cm} 

\begin{table*}[!htbp]
\centering
\footnotesize
\begin{threeparttable}
\caption{Comprehensive ART ANOVA and Post-Hoc Contrasts for Global Spatial Accuracy ($RMSE_{Pos}$).}
\label{tab:anova_rmse_pos}
\begin{tabular*}{\textwidth}{@{\extracolsep{\fill}} l ccccc @{}}
\toprule
\multicolumn{6}{c}{\textbf{Part 1: Omnibus ART ANOVA (Main and Interaction Effects)}} \\
\midrule
\textbf{Effect} & \textbf{\textit{F}-value} & \textbf{\textit{df}} & \textbf{\textit{p}-value} & \multicolumn{2}{c}{\textbf{Effect Size ($\eta_p^2$)}} \\
\midrule
Visualization (UI) & 12.39 & 2, 115 & \textbf{$< .001$} & \multicolumn{2}{c}{0.177} \\
Orientation (Task) & 31.99 & 1, 115 & \textbf{$< .001$} & \multicolumn{2}{c}{0.218} \\
Visualization $\times$ Orientation & 1.05 & 2, 115 & $.354$ & \multicolumn{2}{c}{0.018} \\
\midrule
\multicolumn{6}{c}{\textbf{Part 2: Holm-Corrected Post-Hoc Contrasts}} \\
\midrule
\textbf{Contrast} & \textbf{Estimate} & \textbf{SE} & \textbf{\textit{t}-ratio} & \textbf{\textit{p}\textsubscript{adj}} & \textbf{Effect Size ($d$)} \\
\midrule
\multicolumn{6}{l}{\textit{Main Effect: Visualization}} \\
Gestalt -- Reduced & 11.48 & 5.87 & 1.955 & $.053$ & 0.18 \\
Gestalt -- Tracer    & 29.02 & 5.87 & 4.942 & \textbf{$< .001$} & 0.46 \\
Reduced -- Tracer  & 17.54 & 5.87 & 2.987 & \textbf{$.007$} & 0.28 \\
\midrule
\multicolumn{6}{l}{\textit{Interaction: The Cost of Orientation (Task A [Off] $\rightarrow$ Task B [On] within UIs)}} \\
Gestalt (Off -- On)   & -15.16 & 8.30 & -1.826 & $1.000$ & 0.09 \\
Reduced (Off -- On) & -17.41 & 8.30 & -2.097 & $1.000$ & 0.10 \\
Tracer (Off -- On)    & -29.04 & 8.30 & -3.498 & $.887$ & 0.17 \\
\bottomrule
\end{tabular*}
\begin{tablenotes}[para, flushleft]
\item \textit{Note:} The ART ANOVA utilizes a linear mixed model (LMM) with the participant included as a random effect. Post-hoc contrasts were calculated using Estimated Marginal Means (EMMs) averaged over orientation levels. For the ANOVA, $\eta_p^2$ represents partial eta-squared. For contrasts, $d$ represents Cohen's d approximation.
\end{tablenotes}
\end{threeparttable}
\end{table*}

\begin{table*}[!htbp]
\centering
\footnotesize
\begin{threeparttable}
\caption{Comprehensive ART ANOVA and Post-Hoc Contrasts for Global Velocity Control ($RMSE_{Speed}$).}
\label{tab:anova_rmse_speed}
\begin{tabular*}{\textwidth}{@{\extracolsep{\fill}} l ccccc @{}}
\toprule
\multicolumn{6}{c}{\textbf{Part 1: Omnibus ART ANOVA (Main and Interaction Effects)}} \\
\midrule
\textbf{Effect} & \textbf{\textit{F}-value} & \textbf{\textit{df}} & \textbf{\textit{p}-value} & \multicolumn{2}{c}{\textbf{Effect Size ($\eta_p^2$)}} \\
\midrule
Visualization (UI) & 67.89 & 2, 115 & \textbf{$< .001$} & \multicolumn{2}{c}{0.541} \\
Orientation (Task) & 1.62 & 1, 115 & $.205$ & \multicolumn{2}{c}{0.014} \\
Visualization $\times$ Orientation & 4.03 & 2, 115 & \textbf{$.021$} & \multicolumn{2}{c}{0.065} \\
\midrule
\multicolumn{6}{c}{\textbf{Part 2: Holm-Corrected Post-Hoc Contrasts}} \\
\midrule
\textbf{Contrast} & \textbf{Estimate} & \textbf{SE} & \textbf{\textit{t}-ratio} & \textbf{\textit{p}\textsubscript{adj}} & \textbf{Effect Size ($d$)} \\
\midrule
\multicolumn{6}{l}{\textit{Main Effect: Visualization}} \\
Gestalt -- Reduced & -5.19 & 10.50 & -0.494 & $.263$ & 0.05 \\
Gestalt -- Tracer    & 105.21 & 10.50 & 10.024 & \textbf{$< .001$} & 0.93 \\
Reduced -- Tracer  & 110.39 & 10.50 & 9.481 & \textbf{$< .001$} & 0.88 \\
\midrule
\multicolumn{6}{l}{\textit{Interaction: The Cost of Orientation (Task A [Off] $\rightarrow$ Task B [On] within UIs)}} \\
Gestalt (Off -- On)   & -11.75 & 10.49 & -1.121 & $.472$ & 0.18 \\
Reduced (Off -- On) & -8.17 & 10.49 & -0.778 & $.472$ & 0.12 \\
Tracer (Off -- On)    & -37.04 & 10.49 & -3.529 & \textbf{$.002$} & 0.55 \\
\bottomrule
\end{tabular*}
\begin{tablenotes}[para, flushleft]
\item \textit{Note:} The ART ANOVA utilizes a linear mixed model (LMM) with the participant included as a random effect. Post-hoc contrasts were calculated using Estimated Marginal Means (EMMs) averaged over orientation levels. For the ANOVA, $\eta_p^2$ represents partial eta-squared. For contrasts, $d$ represents Cohen's d approximation.
\end{tablenotes}
\end{threeparttable}
\end{table*}

\begin{table*}[!htbp]
\centering
\footnotesize
\begin{threeparttable}
\caption{Comprehensive ART ANOVA and Post-Hoc Contrasts for Global Orientation Compliance ($RMSE_{Ori}$). \textit{Note: Analysis includes only Task B (Orientation On) trials.}}
\label{tab:anova_rmse_ori}
\begin{tabular*}{\textwidth}{@{\extracolsep{\fill}} l ccccc @{}}
\toprule
\multicolumn{6}{c}{\textbf{Part 1: Omnibus ART ANOVA (Main Effect)}} \\
\midrule
\textbf{Effect} & \textbf{\textit{F}-value} & \textbf{\textit{df}} & \textbf{\textit{p}-value} & \multicolumn{2}{c}{\textbf{Effect Size ($\eta_p^2$)}} \\
\midrule
Visualization (UI) & 13.59 & 2, 58 & \textbf{$< .001$} & \multicolumn{2}{c}{0.319} \\
\midrule
\multicolumn{6}{c}{\textbf{Part 2: Holm-Corrected Post-Hoc Contrasts}} \\
\midrule
\textbf{Contrast} & \textbf{Estimate} & \textbf{SE} & \textbf{\textit{t}-ratio} & \textbf{\textit{p}\textsubscript{adj}} & \textbf{Effect Size ($d$)} \\
\midrule
Gestalt -- Reduced & -11.64 & 25.87 & -0.450 & $.457$ & 0.15 \\
Gestalt -- Tracer    & 65.25 & 25.87 & 2.522 & \textbf{$< .001$} & 0.82 \\
Reduced -- Tracer  & 76.89 & 25.87 & 2.972 & \textbf{$< .001$} & 0.67 \\
\bottomrule
\end{tabular*}
\begin{tablenotes}[para, flushleft]
\item \textit{Note:} The ART ANOVA utilizes a linear mixed model (LMM) with the participant included as a random effect. Post-hoc contrasts were calculated using Estimated Marginal Means (EMMs) averaged over orientation levels. For the ANOVA, $\eta_p^2$ represents partial eta-squared. For contrasts, $d$ represents Cohen's d approximation.
\end{tablenotes}
\end{threeparttable}
\end{table*}

\clearpage

\subsection*{S4. Directional Error Decomposition (Planar vs. Depth)}

To further investigate the spatial deviations, the total positional error was decomposed into a planar component parallel to the user's viewing plane ($RMSE_{XZ}$) and a depth component along the viewing axis ($RMSE_Y$). The descriptive statistics for these directional components are provided in Table \ref{tab:directional_decomposition}. The corresponding ART ANOVA results evaluating the main and interaction effects on both dimensions are detailed in Table \ref{tab:anova_directional}.

\vspace{1.0cm}

\begin{table*}[!htbp]
\centering
\footnotesize
\begin{threeparttable}
\caption{Directional Decomposition of the Spatial Error into Planar ($RMSE_{XZ}$) and Depth ($RMSE_Y$) Components.}
\label{tab:directional_decomposition}
\begin{tabular*}{\textwidth}{@{\extracolsep{\fill}} ll cccc @{}}
\toprule
\textbf{Visualization} & \textbf{Orientation} & $\mathbf{RMSE_{XZ}}$ \textbf{(mm)} & $\mathbf{RMSE_Y}$ \textbf{(mm)} & \textbf{\% in XZ Plane} & \textbf{\% in Y Depth} \\
\midrule
Gestalt   & Off & 3.28 & 3.53 & 48.1\% & 51.9\% \\
Gestalt   & On  & 4.04 & 4.51 & 47.3\% & 52.7\% \\
Reduced   & Off & 3.13 & 3.89 & 44.6\% & 55.4\% \\
Reduced   & On  & 3.79 & 4.40 & 46.3\% & 53.7\% \\
Tracer    & Off & 2.76 & 2.98 & 48.1\% & 51.9\% \\
Tracer    & On  & 3.49 & 3.69 & 48.6\% & 51.4\% \\
\bottomrule
\end{tabular*}
\begin{tablenotes}[para, flushleft]
\item \textit{Note:} ANOVA is performed on absolute error magnitudes; percentages are provided for descriptive visualization of error distribution.
\end{tablenotes}
\end{threeparttable}
\end{table*}

\begin{table*}[!htbp]
\centering
\footnotesize
\begin{threeparttable}
\caption{Comprehensive ART ANOVA and Post-Hoc Contrasts for Directional Error Decomposition.}
\label{tab:anova_directional}
\begin{tabular*}{\textwidth}{@{\extracolsep{\fill}} l cccc | cccc @{}}
\toprule
\multicolumn{9}{c}{\textbf{Part 1: Omnibus ART ANOVA (Main and Interaction Effects)}} \\
\midrule
& \multicolumn{4}{c|}{\textbf{Planar Error ($RMSE_{XZ}$)}} & \multicolumn{4}{c}{\textbf{Depth Error ($RMSE_Y$)}} \\
\cmidrule(lr){2-5} \cmidrule(l){6-9}
\textbf{Effect} & \textbf{\textit{F}-value} & \textbf{\textit{df}} & \textbf{\textit{p}-value} & \textbf{$\eta_p^2$} & \textbf{\textit{F}-value} & \textbf{\textit{df}} & \textbf{\textit{p}-value} & \textbf{$\eta_p^2$} \\
\midrule
Visualization (UI) & 14.67 & 2, 139 & \textbf{$< .001$} & 0.174 & 12.88 & 2, 139 & \textbf{$< .001$} & 0.156 \\
Orientation (Task) & 48.71 & 1, 141 & \textbf{$< .001$} & 0.257 & 20.35 & 1, 141 & \textbf{$< .001$} & 0.126 \\
UI $\times$ Orientation & 0.70 & 2, 139 & $.496$ & 0.010 & 1.21 & 2, 139 & $.301$ & 0.017 \\
\midrule
\multicolumn{9}{c}{\textbf{Part 2: Holm-Corrected Post-Hoc Contrasts}} \\
\midrule
& \multicolumn{4}{c|}{\textbf{Planar Error ($RMSE_{XZ}$)}} & \multicolumn{4}{c}{\textbf{Depth Error ($RMSE_Y$)}} \\
\cmidrule(lr){2-5} \cmidrule(l){6-9}
\textbf{Contrast} & \textbf{Estimate} & \textbf{\textit{t}-ratio} & \textbf{\textit{p}\textsubscript{adj}} & \textbf{$d$} & \textbf{Estimate} & \textbf{\textit{t}-ratio} & \textbf{\textit{p}\textsubscript{adj}} & \textbf{$d$} \\
\midrule
\multicolumn{9}{l}{\textit{Main Effect: Visualization (Averaged over Orientation)}} \\
Gestalt -- Reduced & 16.39 & 2.522 & \textbf{$.013$} & 0.21 & 4.57  & 0.664 & $.508$ & 0.06 \\
Gestalt -- Tracer    & 35.17 & 5.413 & \textbf{$< .001$} & 0.46 & 32.31 & 4.689 & \textbf{$< .001$} & 0.40 \\
Reduced -- Tracer  & 18.78 & 2.891 & \textbf{$.009$} & 0.24 & 27.73 & 4.025 & \textbf{$< .001$} & 0.34 \\
\bottomrule
\end{tabular*}
\begin{tablenotes}[para, flushleft]
\item \textit{Note:} Post-hoc contrasts were calculated using Estimated Marginal Means (EMMs) averaged over orientation levels. 
\end{tablenotes}
\end{threeparttable}
\end{table*}

\clearpage

\subsection*{S5. Subjective Workload, Usability, and Order Effects}

\begin{figure*}[!htbp]
    \centering
    \includegraphics[width=\textwidth]{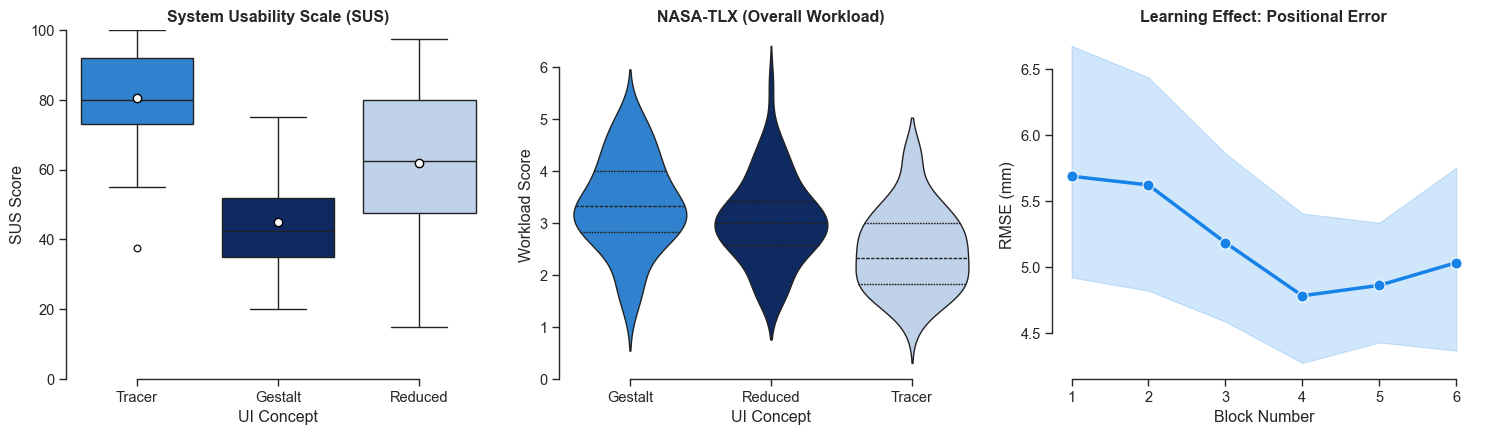}
    \caption{Subjective and objective results of the user study. (Left) System Usability Scale (SUS) scores across the three UI concepts. (Middle) Overall perceived workload measured by NASA-TLX. Boxplot and violin distributions denote quartiles, with white circles indicating the sample means. (Right) The learning effect illustrated by the mean positional error (RMSE) in centimeters over six consecutive trial blocks.}
    \label{fig:results_overview}
\end{figure*}

Figure \ref{fig:results_overview} illustrates the subjective evaluations (SUS, NASA-TLX) alongside the longitudinal block-order effects. The corresponding inferential statistics, detailed in the following tables, utilize linear mixed models (LMM) with the participant included as a random effect. For subjective metrics, the independent variables are the \textit{Visualization} concept and the \textit{Orientation} constraint. For order effects, the independent variable is the trial \textit{Block Number} (1 through 6).

\vspace{1cm} 

\begin{table*}[!htbp]
\centering
\footnotesize
\begin{threeparttable}
\caption{Detailed Item-Level Descriptive Statistics for the System Usability Scale (SUS). Values are formatted as Mean (Standard Deviation) on a 5-point Likert scale, alongside the aggregated total score (Scale 0--100).}
\label{tab:desc_sus}
\begin{tabular*}{\textwidth}{@{\extracolsep{\fill}} l ccc @{}}
\toprule
\textbf{SUS Item} & \textbf{Gestalt (V2)} & \textbf{Reduced (V3)} & \textbf{Tracer (V1)} \\
\midrule
Q01: I think that I would like to use this visualization frequently. & 2.50 (1.20) & 3.30 (1.39) & 4.13 (0.94) \\
Q02: I found the visualization unnecessarily complex. & 3.97 (0.72) & 2.13 (0.94) & 1.63 (0.85) \\
Q03: I thought the visualization was easy to use. & 2.37 (0.93) & 3.23 (1.10) & 4.33 (0.80) \\
Q04: I think that I would need support to be able to use this visualization. & 3.23 (1.28) & 2.63 (1.25) & 1.70 (0.70) \\
Q05: I found the functions in this visualization were well integrated. & 3.40 (0.97) & 3.60 (1.07) & 4.07 (0.78) \\
Q06: I thought there was too much inconsistency in this visualization. & 2.23 (0.82) & 2.23 (0.97) & 2.00 (0.79) \\
Q07: I would imagine that most people would learn to use this visualization very quickly. & 3.03 (0.89) & 3.73 (1.31) & 4.50 (0.68) \\
Q08: I found the visualization very cumbersome to use. & 3.23 (1.17) & 2.50 (1.20) & 1.63 (0.72) \\
Q09: I felt very confident using this visualization. & 2.53 (0.97) & 3.07 (1.31) & 3.83 (0.95) \\
Q10: I needed to learn a lot of things before I could get going with this visualization. & 3.23 (1.07) & 2.63 (1.25) & 1.67 (0.76) \\
\midrule
\textbf{Total SUS Score (Scale 0--100)} & \textbf{44.83 (14.83)} & \textbf{62.00 (21.45)} & \textbf{80.58 (14.69)} \\
\bottomrule
\end{tabular*}
\end{threeparttable}
\end{table*}

\begin{table*}[!htbp]
\renewcommand{\arraystretch}{1.2}
\centering
\footnotesize
\begin{threeparttable}
\caption{Survey Items of the NASA Task Load Index (NASA-TLX) as administered in the study (translated from German). All items were rated on a 5-point Likert scale.}
\label{tab:tlx_items}
\begin{tabularx}{\textwidth}{@{\extracolsep{\fill}} l X @{}}
\toprule
\textbf{NASA-TLX Dimension} & \textbf{English Translation of the Administered Items} \\
\midrule
Mental Demand   & How much mental effort and concentration was required to perform the task with this visualization? \\
Physical Demand & How much physical activity was required to perform the task? (e.g., arm movement, holding work). \\
Temporal Demand & How much time pressure did you feel while performing the task? (Was the pace leisurely or frantic?) \\
Effort          & How hard did you have to work (mentally and physically) to accomplish your level of performance with this visualization? \\
Frustration     & How insecure, discouraged, or annoyed (versus secure, satisfied, relaxed) did you feel during the task? \\
Performance     & How successful do you think you were in accomplishing the goals of the task (position, speed, orientation)? \\
\bottomrule
\end{tabularx}
\end{threeparttable}
\end{table*}

\begin{table*}[!htbp]
\centering
\footnotesize
\begin{threeparttable}
\caption{Descriptive Statistics for NASA-TLX Subscales and Overall Workload (Scale 1--7). Values are formatted as Mean (Standard Deviation).}
\label{tab:desc_tlx}
\begin{tabular*}{\textwidth}{@{\extracolsep{\fill}} ll ccccccc @{}}
\toprule
\textbf{Visualization} & \textbf{Orientation} & \textbf{Overall} & \textbf{Mental Load} & \textbf{Physical} & \textbf{Temporal} & \textbf{Effort} & \textbf{Frustration} & \textbf{Success} \\
\midrule
\multirow{2}{*}{Gestalt}   & Off & 2.98 (0.81) & 3.47 (1.31) & 2.63 (1.35) & 2.97 (1.27) & 3.43 (1.43) & 2.60 (1.25) & 3.23 (1.07) \\
                           & On  & 3.74 (0.82) & 5.28 (1.36) & 3.83 (1.42) & 3.55 (1.35) & 4.79 (1.57) & 3.45 (1.50) & 4.48 (1.38) \\
\midrule
\multirow{2}{*}{Reduced}   & Off & 2.82 (0.61) & 3.03 (1.19) & 2.57 (0.97) & 3.20 (1.45) & 2.93 (1.11) & 2.90 (1.45) & 3.70 (1.42) \\
                           & On  & 3.30 (0.98) & 4.21 (1.63) & 3.34 (1.40) & 3.31 (1.58) & 3.86 (1.46) & 3.52 (1.60) & 4.45 (1.30) \\
\midrule
\multirow{2}{*}{Tracer}    & Off & 2.19 (0.70) & 1.97 (1.10) & 2.03 (0.93) & 1.97 (1.16) & 1.97 (0.93) & 1.77 (1.07) & 2.53 (1.28) \\
                           & On  & 2.79 (0.77) & 2.52 (1.18) & 2.86 (1.13) & 3.45 (1.43) & 2.86 (1.13) & 2.34 (1.23) & 3.28 (1.03) \\
\bottomrule
\end{tabular*}
\end{threeparttable}
\end{table*}

\begin{table*}[!htbp]
\centering
\footnotesize
\begin{threeparttable}
\caption{Comprehensive ART ANOVA and Post-Hoc Contrasts for the System Usability Scale (SUS). \textit{Note: Higher scores indicate better perceived usability.}}
\label{tab:anova_sus}
\begin{tabular*}{\textwidth}{@{\extracolsep{\fill}} l cccc @{}}
\toprule
\multicolumn{5}{c}{\textbf{Part 1: Omnibus ART ANOVA (Main Effect)}} \\
\midrule
\textbf{Effect} & \textbf{\textit{F}-value} & \textbf{\textit{df}} & \textbf{\textit{p}-value} & \textbf{Effect Size ($\eta_p^2$)} \\
\midrule
Visualization (UI) & 33.36 & 2, 58 & \textbf{$< .001$} & 0.535 \\
\midrule
\multicolumn{5}{c}{\textbf{Part 2: Holm-Corrected Post-Hoc Contrasts}} \\
\midrule
\textbf{Contrast} & \textbf{Estimate} & \textbf{\textit{t}-ratio} & \textbf{\textit{p}\textsubscript{adj}} & \textbf{Effect Size ($d$)} \\
\midrule
Gestalt -- Reduced & -20.37 & -3.972 & \textbf{$.0002$} & 0.52 \\
Gestalt -- Tracer    & -41.88 & -8.168 & \textbf{$< .001$} & 1.07 \\
Reduced -- Tracer  & -21.52 & -4.196 & \textbf{$.0002$} & 0.55 \\
\bottomrule
\end{tabular*}
\begin{tablenotes}[para, flushleft]
\item \textit{Note:} The ART ANOVA utilizes a linear mixed model (LMM) with the participant as a random effect. Unlike task-performance metrics, SUS was administered as a global post-test assessment. Consequently, 'Orientation' is not a factor in this model.
\end{tablenotes}
\end{threeparttable}
\end{table*}

\begin{table*}[!htbp]
\centering
\footnotesize
\begin{threeparttable}
\caption{Comprehensive ART ANOVA and Post-Hoc Contrasts for NASA-TLX Primary Metrics (Overall Workload and Mental Demand).}
\label{tab:anova_tlx_main}
\begin{tabular*}{\textwidth}{@{\extracolsep{\fill}} l cccc | cccc @{}}
\toprule
\multicolumn{9}{c}{\textbf{Part 1: Omnibus ART ANOVA (Main and Interaction Effects)}} \\
\midrule
& \multicolumn{4}{c|}{\textbf{Overall Workload}} & \multicolumn{4}{c}{\textbf{Mental Load}} \\
\cmidrule(lr){2-5} \cmidrule(l){6-9}
\textbf{Effect} & \textbf{\textit{F}-value} & \textbf{\textit{df}} & \textbf{\textit{p}-value} & \textbf{$\eta_p^2$} & \textbf{\textit{F}-value} & \textbf{\textit{df}} & \textbf{\textit{p}-value} & \textbf{$\eta_p^2$} \\
\midrule
Visualization (UI) & 36.49 & 2, 142 & \textbf{$< .001$} & 0.339 & 58.89 & 2, 142 & \textbf{$< .001$} & 0.453 \\
Orientation (Task) & 50.15 & 1, 143 & \textbf{$< .001$} & 0.259 & 42.76 & 1, 143 & \textbf{$< .001$} & 0.229 \\
UI $\times$ Orientation & 1.45 & 2, 142 & $.238$ & 0.020 & 5.97 & 2, 142 & \textbf{$.003$} & 0.077 \\
\midrule
\multicolumn{9}{c}{\textbf{Part 2: Holm-Corrected Post-Hoc Contrasts}} \\
\midrule
& \multicolumn{4}{c|}{\textbf{Overall Workload}} & \multicolumn{4}{c}{\textbf{Mental Load}} \\
\cmidrule(lr){2-5} \cmidrule(l){6-9}
\textbf{Contrast} & \textbf{Estimate} & \textbf{\textit{t}-ratio} & \textbf{\textit{p}\textsubscript{adj}} & \textbf{$d$} & \textbf{Estimate} & \textbf{\textit{t}-ratio} & \textbf{\textit{p}\textsubscript{adj}} & \textbf{$d$} \\
\midrule
\multicolumn{9}{l}{\textit{Main Effect: Visualization (Averaged over Orientation)}} \\
Gestalt -- Reduced & 17.44 & 2.710 & \textbf{$.007$} & 0.23 & 25.86 & 3.880 & \textbf{$.0002$} & 0.33 \\
Gestalt -- Tracer    & 53.89 & 8.371 & \textbf{$< .001$} & 0.70 & 71.42 & 10.717 & \textbf{$< .001$} & 0.90 \\
Reduced -- Tracer  & 36.45 & 5.661 & \textbf{$< .001$} & 0.48 & 45.56 & 6.837 & \textbf{$< .001$} & 0.57 \\
\midrule
\multicolumn{9}{l}{\textit{Interaction: The Cognitive Cost of Orientation Constraints (Off -- On within UIs)}} \\
Gestalt (Off -- On)   & -39.41 & -4.586 & \textbf{$< .001$} & 0.38 & -48.43 & -5.520 & \textbf{$< .001$} & 0.46 \\
Reduced (Off -- On) & -24.24 & -2.820 & \textbf{$.028$} & 0.24 & -32.35 & -3.688 & \textbf{$.002$} & 0.31 \\
Tracer (Off -- On)    & -34.86 & -4.056 & \textbf{$.001$} & 0.34 & -17.14 & -1.953 & $.142$ & 0.16 \\
\bottomrule
\end{tabular*}
\begin{tablenotes}[para, flushleft]
\item \textit{Note:} Interaction effects are explicitly reported for Overall Workload and Mental Demand, as these were defined a priori as the primary cognitive metrics of interest.
\end{tablenotes}
\end{threeparttable}
\end{table*}

\begin{table*}[!htbp]
\centering
\footnotesize
\begin{threeparttable}
\caption{Comprehensive ART ANOVA and Post-Hoc Contrasts for NASA-TLX Secondary Subscales. \textit{Note: Lower scores indicate lower perceived workload.}}
\label{tab:anova_tlx_subscales}
\begin{tabular*}{\textwidth}{@{\extracolsep{\fill}} l | ccc | ccc | ccc @{}}
\toprule
\multicolumn{10}{c}{\textbf{Part 1: Omnibus ART ANOVA (Main and Interaction Effects)}} \\
\midrule
& \multicolumn{3}{c|}{\textbf{Visualization (UI)}} & \multicolumn{3}{c|}{\textbf{Orientation (Task)}} & \multicolumn{3}{c}{\textbf{UI $\times$ Orientation}} \\
\cmidrule(lr){2-4} \cmidrule(lr){5-7} \cmidrule(l){8-10}
\textbf{NASA-TLX Subscale} & \textbf{\textit{F}-value} & \textbf{\textit{p}-value} & \textbf{$\eta_p^2$} & \textbf{\textit{F}-value} & \textbf{\textit{p}-value} & \textbf{$\eta_p^2$} & \textbf{\textit{F}-value} & \textbf{\textit{p}-value} & \textbf{$\eta_p^2$} \\
\midrule
Physical Demand & 9.88 & \textbf{$< .001$} & 0.122 & 47.06 & \textbf{$< .001$} & 0.247 & 1.36 & $.261$ & 0.019 \\
Temporal Demand & 4.49 & \textbf{$.013$} & 0.060 & 26.15 & \textbf{$< .001$} & 0.154 & 4.90 & \textbf{$.009$} & 0.065 \\
Effort          & 38.63 & \textbf{$< .001$} & 0.352 & 60.26 & \textbf{$< .001$} & 0.295 & 2.33 & $.101$ & 0.032 \\
Frustration     & 27.58 & \textbf{$< .001$} & 0.280 & 14.09 & \textbf{$< .001$} & 0.090 & 0.19 & $.829$ & 0.003 \\
Success & 21.65 & \textbf{$< .001$} & 0.234 & 35.61 & \textbf{$< .001$} & 0.199 & 0.88 & $.416$ & 0.012 \\
\midrule
\multicolumn{10}{c}{\textbf{Part 2: Holm-Corrected Post-Hoc Contrasts (Main Effect of Visualization)}} \\
\midrule
\textbf{Contrast (per Subscale)} & \multicolumn{2}{c}{\textbf{Estimate}} & \multicolumn{2}{c}{\textbf{\textit{t}-ratio}} & \multicolumn{2}{c}{\textbf{\textit{p}\textsubscript{adj}}} & \multicolumn{3}{c}{\textbf{Effect Size ($d$)}} \\
\midrule
\multicolumn{10}{l}{\textit{Physical Demand}} \\
\quad Gestalt -- Reduced & \multicolumn{2}{c}{14.12} & \multicolumn{2}{c}{1.991} & \multicolumn{2}{c}{\textbf{$.048$}} & \multicolumn{3}{c}{0.16} \\
\quad Gestalt -- Tracer    & \multicolumn{2}{c}{31.47} & \multicolumn{2}{c}{4.437} & \multicolumn{2}{c}{\textbf{$< .001$}} & \multicolumn{3}{c}{0.37} \\
\quad Reduced -- Tracer  & \multicolumn{2}{c}{17.35} & \multicolumn{2}{c}{2.446} & \multicolumn{2}{c}{\textbf{$.031$}} & \multicolumn{3}{c}{0.20} \\
\midrule
\multicolumn{10}{l}{\textit{Temporal Demand}} \\
\quad Gestalt -- Reduced & \multicolumn{2}{c}{2.22}  & \multicolumn{2}{c}{0.290} & \multicolumn{2}{c}{$.772$} & \multicolumn{3}{c}{0.02} \\
\quad Gestalt -- Tracer    & \multicolumn{2}{c}{20.95} & \multicolumn{2}{c}{2.729} & \multicolumn{2}{c}{\textbf{$.021$}} & \multicolumn{3}{c}{0.22} \\
\quad Reduced -- Tracer  & \multicolumn{2}{c}{18.72} & \multicolumn{2}{c}{2.439} & \multicolumn{2}{c}{\textbf{$.031$}} & \multicolumn{3}{c}{0.20} \\
\midrule
\multicolumn{10}{l}{\textit{Effort}} \\
\quad Gestalt -- Reduced & \multicolumn{2}{c}{23.58} & \multicolumn{2}{c}{3.355} & \multicolumn{2}{c}{\textbf{$.001$}} & \multicolumn{3}{c}{0.28} \\
\quad Gestalt -- Tracer    & \multicolumn{2}{c}{61.24} & \multicolumn{2}{c}{8.713} & \multicolumn{2}{c}{\textbf{$< .001$}} & \multicolumn{3}{c}{0.73} \\
\quad Reduced -- Tracer  & \multicolumn{2}{c}{37.66} & \multicolumn{2}{c}{5.358} & \multicolumn{2}{c}{\textbf{$< .001$}} & \multicolumn{3}{c}{0.45} \\
\midrule
\multicolumn{10}{l}{\textit{Frustration}} \\
\quad Gestalt -- Reduced & \multicolumn{2}{c}{-4.30} & \multicolumn{2}{c}{-0.674} & \multicolumn{2}{c}{$.501$} & \multicolumn{3}{c}{0.05} \\
\quad Gestalt -- Tracer    & \multicolumn{2}{c}{38.68} & \multicolumn{2}{c}{6.069} & \multicolumn{2}{c}{\textbf{$< .001$}} & \multicolumn{3}{c}{0.50} \\
\quad Reduced -- Tracer  & \multicolumn{2}{c}{42.98} & \multicolumn{2}{c}{6.743} & \multicolumn{2}{c}{\textbf{$< .001$}} & \multicolumn{3}{c}{0.56} \\
\midrule
\multicolumn{10}{l}{\textit{Success}} \\
\quad Gestalt -- Reduced & \multicolumn{2}{c}{-12.23} & \multicolumn{2}{c}{-1.693} & \multicolumn{2}{c}{$.092$} & \multicolumn{3}{c}{0.14} \\
\quad Gestalt -- Tracer    & \multicolumn{2}{c}{33.66} & \multicolumn{2}{c}{4.660} & \multicolumn{2}{c}{\textbf{$< .001$}} & \multicolumn{3}{c}{0.39} \\
\quad Reduced -- Tracer  & \multicolumn{2}{c}{45.88} & \multicolumn{2}{c}{6.353} & \multicolumn{2}{c}{\textbf{$< .001$}} & \multicolumn{3}{c}{0.53} \\
\bottomrule
\end{tabular*}
\begin{tablenotes}[para, flushleft]
\item \textit{Note:} Degrees of freedom are ($2, 142$) for Visualization and the Interaction, and ($1, 143$) for Orientation. Post-hoc contrasts in Part 2 isolate the main effect of Visualization, averaged across orientation constraints.
\end{tablenotes}
\end{threeparttable}
\end{table*}

\begin{table*}[!htbp]
\centering
\footnotesize
\begin{threeparttable}
\caption{Comprehensive ART ANOVA and Post-Hoc Contrasts for Block Order Effects on Spatial Error ($RMSE_{Pos}$). \textit{Note: The lack of significance confirms the successful mitigation of learning and fatigue effects.}}
\label{tab:anova_block_order}
\begin{tabular*}{\textwidth}{@{\extracolsep{\fill}} l cccc @{}}
\toprule
\multicolumn{5}{c}{\textbf{Part 1: Omnibus ART ANOVA (Main Effect)}} \\
\midrule
\textbf{Effect} & \textbf{\textit{F}-value} & \textbf{\textit{df}} & \textbf{\textit{p}-value} & \textbf{Effect Size ($\eta_p^2$)} \\
\midrule
Block Number & 1.65 & 5, 140 & $.152$ & 0.056 \\
\midrule
\multicolumn{5}{c}{\textbf{Part 2: Holm-Corrected Post-Hoc Contrasts}} \\
\midrule
\textbf{Contrast} & \textbf{Estimate} & \textbf{\textit{t}-ratio} & \textbf{\textit{p}\textsubscript{adj}} & \textbf{Effect Size ($d$)} \\
\midrule
Block 1 -- Block 2 & 6.17  & 0.591 & $1.000$ & 0.05 \\
Block 1 -- Block 3 & 14.03 & 1.344 & $1.000$ & 0.11 \\
Block 1 -- Block 4 & 24.67 & 2.414 & $.256$  & 0.20 \\
Block 1 -- Block 5 & 19.37 & 1.896 & $.799$  & 0.16 \\
Block 1 -- Block 6 & 20.03 & 1.919 & $.799$  & 0.16 \\
Block 2 -- Block 3 & 7.86  & 0.743 & $1.000$ & 0.06 \\
Block 2 -- Block 4 & 18.50 & 1.772 & $.943$  & 0.15 \\
Block 2 -- Block 5 & 13.20 & 1.264 & $1.000$ & 0.11 \\
Block 2 -- Block 6 & 13.86 & 1.310 & $1.000$ & 0.11 \\
Block 3 -- Block 4 & 10.64 & 1.019 & $1.000$ & 0.09 \\
Block 3 -- Block 5 & 5.34  & 0.511 & $1.000$ & 0.04 \\
Block 3 -- Block 6 & 6.00  & 0.567 & $1.000$ & 0.05 \\
Block 4 -- Block 5 & -5.30 & -0.519 & $1.000$ & 0.04 \\
Block 4 -- Block 6 & -4.64 & -0.444 & $1.000$ & 0.04 \\
Block 5 -- Block 6 & 0.66  & 0.063 & $1.000$ & 0.01 \\
\bottomrule
\end{tabular*}
\end{threeparttable}
\end{table*}

\clearpage

\subsection*{S6. Qualitative Analysis}

Table \ref{tab:detailed_qualitative_feedback} provides a thematic breakdown of the free-text user feedback, contextualizing the statistical data. 
\vspace{1cm} 

\begin{table*}[!htbp]
\centering
\footnotesize
\begin{threeparttable}
\caption{Detailed Qualitative User Feedback and Thematic Analysis (N=30)}
\label{tab:detailed_qualitative_feedback}
\begin{tabularx}{\textwidth}{l l c X}
\toprule
\textbf{Variant} & \textbf{Identified Theme} & \textbf{$n$} & \textbf{Scientific Interpretation \& Honest User Feedback} \\
\midrule

\textbf{V1 (Ghost)} 
& \textbf{[Strength]} Unified Attentional Focus & 12 & 
Bundling the target position and speed into a single focal object minimizes cognitive load. Users rely on mimicry rather than mental calculation. \newline
\textit{Quote: "Visualizing the speed using a ghost point created cognitive relief. [...] It is pleasant that you do not have to read any displays, everything is implicit."} \\
\addlinespace
& \textbf{[Limitation]} Visual Occlusion \& Depth & 4 & 
The target marker on the trajectory perfectly overlapping the actual position marker blocks the view. Furthermore, assessing depth from a constant perspective is difficult. \newline
\textit{Quote: "The target marker (white sphere) obscures the marker of the current position. Three-dimensional estimation is difficult with a constant viewing angle."} \\
\addlinespace
& \textbf{[Limitation]} Lack of Anticipation & 2 & 
Users felt surprised by sudden changes in the trajectory's parameters, leading to abrupt tracking errors. \newline
\textit{Quote: "A warning before an upcoming change in speed or angle would be helpful."} \\
\midrule

\textbf{V2 (Gestalt)} 
& \textbf{[Strength]} High-Precision Anchoring & 4 & 
Highly beneficial for deterministic, fine-grained angle corrections. The explicit coordinate system removes the physical need to move the head to check alignment. \newline
\textit{Quote: "The visualization of the pose was the most helpful here, otherwise moving the head is always necessary to correct properly from another perspective."} \\
\addlinespace
& \textbf{[Limitation]} Severe Split Attention & 15 & 
The spatial separation of the tool tip (trajectory focus) and the UI displays (speedometer) forces constant saccades, leading to rapid visual and cognitive overload. \newline
\textit{Quote: "Too many redundant elements. The distances between the tip visualization and the speed were too far apart to keep an eye on both simultaneously."} \\
\addlinespace
& \textbf{[Limitation]} Inclusivity \& Clutter & 3 & 
The sheer volume of lines is overwhelming. Furthermore, color coding decisions were critiqued for lack of accessibility. \newline
\textit{Quote: "Red and green as colors are not inclusive; leads to comprehension problems. Information cannot be processed due to an overabundance of representations."} \\
\midrule

\textbf{V3 (Reduced)} 
& \textbf{[Strength]} Tactile Offloading & 5 & 
Shifting the speed feedback to the haptic channel was perceived as a valid strategy to reduce visual clutter and keep the eyes on the physical trajectory. \\
\addlinespace
& \textbf{[Limitation]} Dimensional Ambiguity & 14 & 
The haptic/visual feedback is strictly binary (error presence) and lacks a directional vector (error magnitude/direction). This ambiguity induces stress and incorrect adjustments. \newline
\textit{Quote: "Vibrating does not allow you to feel whether you are too fast or too slow. Two different vibration patterns would be highly helpful."} \\
\addlinespace
& \textbf{[Strength]} Efficacy of Constraints & 5 & 
Geometric constraints (tolerance arrows/triangles) effectively compensate for AR perspective issues, providing clear error amplitudes regardless of the user's viewing angle. \newline
\textit{Quote: "Position tolerance arrows are very helpful, as they visualize the error amplitude well even from an unfavorable viewing angle."} \\
\midrule

\textbf{Synthesis} 
& Hybrid Design Request & 4 & 
Without prompting, multiple users independently proposed combining the strengths of the variants: Using V1 (Ghost) for pacing and global trajectory, combined with V3 (Arrows) for fine-tuning. \newline
\textit{Quote: "I would like to continue following the ball and maybe bring in the coordinate system [or arrows] so you can fine-tune the orientation."} \\
\bottomrule
\end{tabularx}
\end{threeparttable}
\end{table*}

\end{document}